\documentclass[12pt]{article}
\usepackage{color,amsmath,amssymb,exscale,psfrag,epsfig}
\input epsf
\newcommand{\m}[1]{\marginpar{{\tiny *}} }
\newcommand{\Gslash}{{\not \!\! G}}

\def\bea{\begin{eqnarray}}
\def\eea{\end{eqnarray}}
\newcommand{\mtt}{$M_{t\bar t}$-spectrum }

\begin{document}
\topmargin -1.0cm
\oddsidemargin -0.8cm
\evensidemargin -0.8cm

\hfill
\vspace{20pt}
\begin{center}
{\Large \bf Phenomenology of a light gluon resonance in top-physics at Tevatron and LHC} 
\end{center}

\vspace{15pt}
\begin{center}
{\large{Ezequiel \'Alvarez\footnote{sequi@unsl.edu.ar}$^{,a}$, Leandro Da Rold\footnote{daroldl@cab.cnea.gov.ar}$^{,b}$, Juan Ignacio Sanchez Vietto\footnote{jisanchez@df.uba.ar}$^{,c}$, Alejandro Szynkman\footnote{szynkman@fisica.unlp.edu.ar}$^{,d}$}}

\vspace{20pt}
$^{a}$\textit{CONICET, INFAP and Departamento de F\'{\i}sica, FCFMN, Universidad de San Luis \\
Av. Ej\'ercito de los Andes 950, 5700, San Luis, Argentina}
\\[0.2cm]
$^{b}$\textit{CONICET, Centro At\'omico Bariloche and Instituto Balseiro\\
Av.\ Bustillo 9500, 8400, S.\ C.\ de Bariloche, Argentina }
\\[0.2cm]
$^{c}$\textit{CONICET, IFIBA and Departamento de F\'{\i}sica, FCEyN, Universidad de Buenos Aires \\
Ciudad Universitaria, Pab.1, (1428) Buenos Aires, Argentina}
\\[0.2cm]
$^{d}$\textit{ IFLP, CONICET - Dpto. de F\'{\i}sica, Universidad Nacional de La Plata, C.C. 67, 1900 La Plata, Argentina}
\end{center}

\begin{abstract}
We present a phenomenological analysis of the recent Tevatron results
on the $t\bar t$ forward-backward asymmetry and invariant-mass
spectrum assuming a new contribution from an $s$-channel gluon resonance with a mass in the range from $700$ to $2500$ GeV.  In contrast to
most of the previous works, this analysis shows that for masses below $\sim 1$ TeV
resonant New Physics could accommodate the experimental data.  In
general, we find that axial-like couplings are preferred for light and
top quark couplings, and that only top quark couples strongly to New
Physics.  We find that composite model scenarios arise naturally from
only phenomenological analyses of the experimental results.  We show
that our results are compatible with recent LHC limits in dijet and
$t\bar t$ production, and find some tension for large resonance mass 
$\sim 2.5$ TeV.  We indicate as best observables for discriminating
a relatively light new gluon a better resolution in CDF forward-backward
asymmetry, as well as the $t\bar t$ charge asymmetry and invariant-mass
spectrum at the LHC.
\end{abstract}

\newpage

\setcounter{footnote}{0}

\section{Introduction}
The Standard Model (SM) has accurately explained with great success the physics up to the $\sim 100$ GeV scale. However, there are several hints that indicate that New Physics (NP) could be expected at the TeV scale.  Among the main motivations to look for NP at this scale, we can mention the quadratic divergences of the Higgs mass and the mechanism responsible for spontaneous symmetry breaking.

To explore the TeV-scale physics, physicists have built two colossal
machines, Tevatron and LHC.  Tevatron has not found signals of the Higgs boson, but it has studied as never before the top quark sector. The LHC has begun its search for the Higgs boson, which could take a few years.  In the mean time,
the LHC is becoming the largest {\it top factory} ever built.

Experiments in Tevatron and LHC have a special focus in top quark physics since, from a theoretical point of view, it is expected that the top quark could be a window to the NP \cite{topNP}.  In fact, mainly because its extraordinary large mass and secondly because its poor exploration insofar,  it is expected that the study of top quark production and the analysis of its properties, as couplings, mass and width, may throw some light on how is the NP.  As a matter of fact, CDF and D0 have published several articles \cite{CDF1-Asym,CDF2-Asym,D01-Asym} in the last few years on $t\bar t$ pair-production, where in CDF results the forward-backward asymmetry ($A_{FB}^t$) measurement did not agree with the SM NLO (next-to-leading order) expectation, whereas the cross section ($\sigma_{t\bar t}$) and its invariant-mass ($M_{t\bar t}$) spectrum did agree with the SM NLO expectation.  These results have raised the attention of the theoretical community which has essayed their hypothesis in several and varied publications \cite{ttbar}.  In a recent CDF article \cite{CDF3-Asym} on this issue, it was found that the total $A_{FB}^t$ agrees with the SM expectation, but when studied as a function of the invariant mass of the top pair a large deviation ($\sim 3\sigma$) is found for large invariant mass.  On the other hand, the cross section and invariant-mass spectrum still agree with the SM expectation.  The results in that article, which have also motivated many theoretical publications \cite{ttbar2}, are the motivation for the present work.

CDF has reported in Ref.~\cite{CDF3-Asym} that the forward-backward asymmetry measured in $p\bar p \to t \bar t$ processes deviates from the SM prediction when the invariant mass of the $t\bar t$ pair is above $450$ GeV.  This deviation, when analyzed from the not-unfolded results, consists in 4 bins in the range $450\mbox{ GeV}<M_{t\bar t}<700\mbox{ GeV}$ in which, although with little statistical significance, {\it all} the measurements are {\it above} the SM expected.  When this data is unfolded and computed in only 2 bins for $M_{t\bar t}\gtrless 450$ GeV, then a $3.4\ \sigma$ disagreement is found in the higher invariant mass bin.  

Previous works \cite{ttbar2} have proposed new particles in $t$- and $s$-channel, and non-resonant NP as effective operators to solve this anomaly. Moreover, in all the $s$-channel proposals, the resonance mass is always pushed above $\sim 1.5$ TeV in order to avoid spoiling the invariant-mass spectrum\footnote{During the preparation of this article \cite{talkPlanck2011}, a work analyzing the possibility of a light $s$-channel resonance came out~\cite{granada}. In the region of the parameter space shared with the present article we reach similar conclusions.}.  In this work, motivated by a qualitative analysis of the not-unfolded data on the $A_{FB}^t$ in Ref.~\cite{CDF3-Asym} --which has much better resolution than the unfolded results--, we explore the possibility of having an $s$-channel contribution of a new particle whose mass could be as light as $\sim 700$ GeV.  In fact, the excess in the not-unfolded data in $A_{FB}^t$ may be suggesting the well studied tail of a non-resonant contribution, as well as a light ($\sim 700-1000$ GeV) resonance peak. Henceforth, we have explored the phenomenology of an $s$-channel colour-octet resonance with chiral couplings in a mass range from $700$ GeV to $2500$ GeV.  We have required for the light resonance to be weakly coupled to light quarks in order to avoid spoiling the $t\bar t $ invariant-mass spectrum and the colliders jet phenomenology.  As it is shown in the article, these and other phenomenological requirements give rise naturally to composite model scenarios where only the top quark couples strongly to the NP.  Moreover, we found that axial-like couplings for light and top quarks are preferred in all the studied mass range.  Although we found that larger masses are preferred, these are more likely to be excluded by the dijet phenomenology.

It is worth stressing at this point that, although the motivation for this work came from a qualitative analysis of the above mentioned not-unfolded data, the quantitative analysis has been performed against the unfolded data.

This work is divided as follows.  In the next section we present the effective model to be used in the analysis.  In section \ref{analytic} we perform an analytic study of the process $q\bar q \to t \bar t$, paying special attention to the resonant and interfering terms in order to determine which are the best regions of the parameter space to look for agreement with the CDF observables $A_{FB}^t$ and $M_{t\bar t}$-spectrum.  In section \ref{numeric} we present the results of Monte Carlo $p\bar p\to t\bar t$ simulations, exploring the parameter space of the effective model which satisfies the CDF observables.  Section \ref{lhc} analyzes the favourable points in parameter space found in previous section within the existing constraints and perspectives for LHC.  Section \ref{models} contains an analysis and discussion of the favourable points within the framework of some popular models as warped extra-dimensions and composite models.  Section \ref{conclusions} contains the conclusions and final ideas.

\section{Effective model}
\label{effective}

We consider a model with a light gluon resonance transforming as an octet of SU(3)$_c$ and mass $M\sim$ TeV. Such a scenario is naturally obtained within the framework of composite models and extra dimensions, since they typically contain this class of resonances~\cite{gaugeKK}. The interactions with the SM-quarks are given by:
\begin{equation}\label{gA*}
{\cal L}_{int}\supset \sum_q g_s (f_{q_L} \ \bar q_L \ \Gslash^* q_L + f_{q_R} \ \bar q_R \ \Gslash^* q_R) \ ,
\end{equation}
where $f_{q_{L,R}}$ are the chiral couplings between the gluon resonance and the SM quarks, in units of the strong coupling $g_s$. For simplicity we will consider that the light quarks have the same chiral couplings:
\begin{equation}\label{fq}
f_{u_L}=f_{d_L}=f_{c_L}=f_{s_L}=f_{q_L} \ , \qquad f_{u_R}=f_{d_R}=f_{c_R}=f_{s_R}=f_{q_R} \ ,
\end{equation}
and we will allow different couplings for the third generation:
\begin{equation}\label{ft}
f_{t_L}=f_{b_L} \ , \qquad f_{t_R} \ , \qquad f_{b_R} \ .
\end{equation}

For a more complete description of a large class of effective theories containing a gluon resonance as well as electroweak (EW) and fermionic resonances, look Ref.~\cite{Contino:2006nn} and Refs.~\cite{DaRold:2010as,Alvarez:2010js,Djouadi:2009nb}.

In warped extra-dimensional and composite Higgs models $f_q$ is usually correlated with the Yukawa couplings, leading to $f_{c,s}>f_{u,d}$. However the quark content of the proton is mostly $u$ and $d$ quarks, thus at leading order in our model $f_c, \ f_s$ and $f_b$ only enter in the width of the resonance. As we discuss next, also the recent experimental results from Tevatron prefer $f_{t,b}>f_q$, leading to a resonance width mostly determined by the couplings $f_{t,b}$. In this scenario top-pair creation in proton collisions are not sensitive to small differences between the couplings of the first and second generations, as long as they remain small compared with the couplings of the third generation, justifying our simplified assignment. The anomaly in $A^b_{FB}$ measured at LEP and SLC can be solved if $b_R$ is partially composite, leading to larger couplings for $b_R$ than for the light generations. For this reason we consider $f_{b_R}$ as an independent coupling and allow it to be larger than $f_q$ â~\cite{DaRold:2010as,Alvarez:2010js,Djouadi:2006rk,Djouadi:2011aj}. For $f_{t,b}\gtrsim 1$, the resonance width is comparable to the resonance mass: $\Gamma/M\sim 0.1-1$, therefore we will consider an energy dependent width for the resonance.

\section{Analytic study of the $q\bar q \to t\bar t$ process}
\label{analytic}
Within the framework of the effective model considered in the previous section, we perform an analytic parton-level analysis of both the forward-backward asymmetry and the cross section of $t\bar t$ pair production at Tevatron. The NP contribution to $t \bar t$ production  in Tevatron comes from a gluon resonance exchange in the $s$-channel of the $q\bar q \to t \bar t$ process. Henceforth, in order to obtain a qualitative understanding of the asymmetry and the cross section at Tevatron, we study the $s$-dependent tree-level parton formulae in the center of mass frame of the $t\bar t$ system. In the following section, we refine the analysis turning to a Monte-Carlo simulation of the $p\bar p \to t\bar t$ process and include the SM-NLO result that gives a non-vanishing contribution to the SM expected value of $A^{t}_{FB}$.

When considering $t\bar t$ as the final state, we can write the differential angular cross section in the center of mass reference frame of the $t\bar t$ system as
\begin{equation}
\frac{d\sigma}{d\cos\theta}=
\mathcal{A}^{\mbox{\tiny SM}}+\mathcal{A}^{\mbox{\tiny INT}}+\mathcal{A}^{\mbox{\tiny NPS}} 
\label{diffcross}
\end{equation}
where INT and NPS stand for {\it SM-NP interference} and {\it new physics squared}, respectively.  At tree-level, including only LO QCD and the gluon resonance contributions, the three different terms read
\begin{eqnarray}
\mathcal{A}^{\mbox{\tiny SM}} & = &
 \frac{\pi\beta\alpha_{s}^{2}}{9 s}\left(2-\beta^{2}
 +\left(\beta\cos\theta\right)^{2}\right) \, , \label{aes1} \\
\mathcal{A}^{\mbox{\tiny INT}} & = &
 \frac{2\pi\beta\alpha_{s}^{2}}{9 s}
 \frac{s \left(s-M^{2}\right)}
{\left(s-M^{2}\right)^{2}
+ M^2 \Gamma^2_{G^*}(s)}
v_q v_t \nonumber\\
 & \times & 
 \left\{ \left(2-\beta^{2}\right)+
 2 \frac{a_q a_t}{v_q v_t} \beta\cos\theta+
 \left(\beta\cos\theta\right)^{2}\right\} \, , \label{aes2} \\
\mathcal{A}^{\mbox{\tiny NPS}} & = &
 \frac{\pi\beta\alpha_{s}^{2}}{9 s}
 \frac{s^{2}}{\left(s-M^{2}\right)^{2}
 + M^2 \Gamma^2_{G^*}(s)}(v_q^2+a_q^2)(v_t^2+a_t^2)\nonumber \\
 & \times & 
 \left\{ 1+\frac{v^2_t-a^2_t}{v_t^2+a_t^2}\left(1-\beta^{2}\right)
 +8\frac{v_q a_q v_t a_t} 
 {(v_q^2+a_q^2)(v_t^2+a_t^2)}
 \beta\cos\theta+\left(\beta\cos\theta\right)^{2}\right\} \, .
\label{aes3}
\end{eqnarray} 
The SM contribution corresponds to the $s$-channel gluon exchange diagram, the NPS term is obtained by replacing the gluon by $G^*$ in that diagram, and INT stands for the gluon-$G^*$ interference. The angle $\theta$ is defined by the directions of motion of the top quark and the incoming light quark (up quark, for instance) in the center of mass frame of the top pair system. The coefficient $v_q(a_q)$ denotes the vector(axial) coupling between a quark of the light generations and $G^*$ in units of $g_{s}$, whereas $v_t(a_t)$ is an analogous notation for the top couplings~\footnote{The vector and axial couplings are given in terms of the chiral couplings introduced in Section~\ref{effective} as \\ $v_{q,t}=(f_{{(q,t)}_{R}}+f_{{(q,t)}_{L}})/2$ and $a_{q,t}=(f_{{(q,t)}_{R}}-f_{{(q,t)}_{L}})/2$.}. Finally, $M$ represents the mass of $G^*$, $\sqrt s$ is the energy in the center of mass frame of the $t \bar t$ system and $\beta=\sqrt{1-4m_t^2/s}$ is the velocity of the top in that frame. Since the narrow-width approximation is no longer valid for large couplings, the use of a modified Breit-Wigner function is needed for the $G^*$ propagator in Eqs. (\ref{aes2},\ref{aes3}), where $\Gamma_{G^*}(s)$ is an energy dependent width. 

The forward and backward cross sections are defined as
\bea
\sigma_F\equiv\int_{0}^{1}\frac{d\sigma}{d\cos\theta}d\cos\theta \, ,
\qquad\sigma_B\equiv\int_{-1}^{0}\frac{d\sigma}{d\cos\theta}d\cos\theta \, ,
\eea
and $A^{t}_{FB}$ is given by
\bea
A^{t}_{FB} = \frac{\sigma_F-\sigma_B}{\sigma_F+\sigma_B} \, . 
\eea
Note that just the constant and $\cos^2\theta$ terms contribute to $\sigma_{t\bar t}$, whereas only the $\cos\theta$ term gives a contribution to the numerator of $A^{t}_{FB}$.

If the NP effects in the cross section are small compared to the leading SM contribution, $\sigma^{SM}\gg\sigma^{NP}$ (where $\sigma^{NP}$ contains both INT and NPS contributions), we can approximate $A_{FB}^t$ by 
\bea
A^{t}_{FB} \approx A_{FB}^{t(SM)} + A_{FB}^{t (NP)}\ ,
\label{Asym}
\eea
for each bin in the $A_{FB}^t$ invariant-mass distribution. The numerator of $A_{FB}^{t (NP)}$ includes INT+NPS terms given at tree-level by Eqs.~(\ref{aes2}) and (\ref{aes3}), whereas the denominator is dominated by the SM tree-level contribution given by Eq.~(\ref{aes1}). We use the NLO result for $A_{FB}^{t(SM)}$ in order to take into account the SM contribution to the asymmetry.  We have checked that the condition $\sigma^{SM}\gg\sigma^{NP}$ is satisfied in our simulations.

The main point of this section is to gain a preliminar insight on which parameter space regions are favored by the experimental results. According to this, we analyze Eqs.~(\ref{aes1}-\ref{aes3}) and look for general conditions to be fulfilled by the couplings. In particular, motivated by the not-unfolded data on $A^{t}_{FB}$ (see Fig.~10 in Ref.~\cite{CDF3-Asym}), we pursue to reproduce the peak around 650 GeV as a resonant effect arising from a $G^*$ with $M \gtrsim 700$ GeV and, at the same time, to produce small corrections to the invariant mass distribution in the cross section. As a first observation, we notice that the NPS contributions to $A^{t}_{FB}$ dominate over the INT terms in the resonant region. Since a positive asymmetry is measured in that region, for typical values $\sqrt s\sim 500$ GeV, a first constraint is imposed on the couplings: $v_q a_q v_t a_t > 0$. Subsequently, it is also required $a_q a_t < 0$ in order to assure a positive pre-peak since INT contributions may become competitive in the region $s \lesssim {(M-\Gamma_{G^*})}^2$ where the pre-peak takes place. In addition, these two conditions imply that $v_q v_t < 0$, giving positive and negative contributions to $\sigma_{t \bar t}$ from INT terms in the regions with $s \lesssim {(M-\Gamma_{G^*})}^2$ and $s \gtrsim {(M+\Gamma_{G^*})}^2$, respectively. In order to avoid large distortions in the $\sigma_{t \bar t}$ invariant mass profile and significant deviations from the dijet final state invariant mass distributions, we demand small axial and vector couplings for light quarks --we satisfy this condition by assuming small chiral couplings--. On the other side, we set large top chiral couplings for two reasons: (i)  to make the $G^*$ effects detectable and (ii) to explain the $A^t_{FB}$ shape by means of the presence of a relatively broad resonance --we will see below that a narrow resonance does not lead to an agreement with the experimental results.  Besides, NPS contributions to $\sigma_{t \bar t}$ are always positive for any choice of couplings as we can read from Eq.~(\ref{aes3}). Therefore, we expect the cross section to increase in the resonant region. Thus, we see that a delicate interplay between the size of $G^*$ width and its couplings to quarks for fixed values of $M$ dictates the experimentally allowed regions in the parameter space. It is worth to notice that a clear signal of the presence of a light resonance would be to detect  an excess of events in the invariant mass distribution of $A^{t}_{FB}$ and $\sigma_{t \bar t}$ for $s$ values below the resonant peak and a defect above. 

\begin{figure}[!htb]
\psfrag{x}{\scriptsize $M_{t \bar t}$ [GeV]}
\psfrag{y}{\scriptsize $ \sim \frac{d\sigma_{t\bar t}}{dM_{t \bar t}}$}
\begin{center}
\hspace*{-1.4cm}
\begin{minipage}[b]{0.39\linewidth}
\begin{center}
\includegraphics[width=1\textwidth]{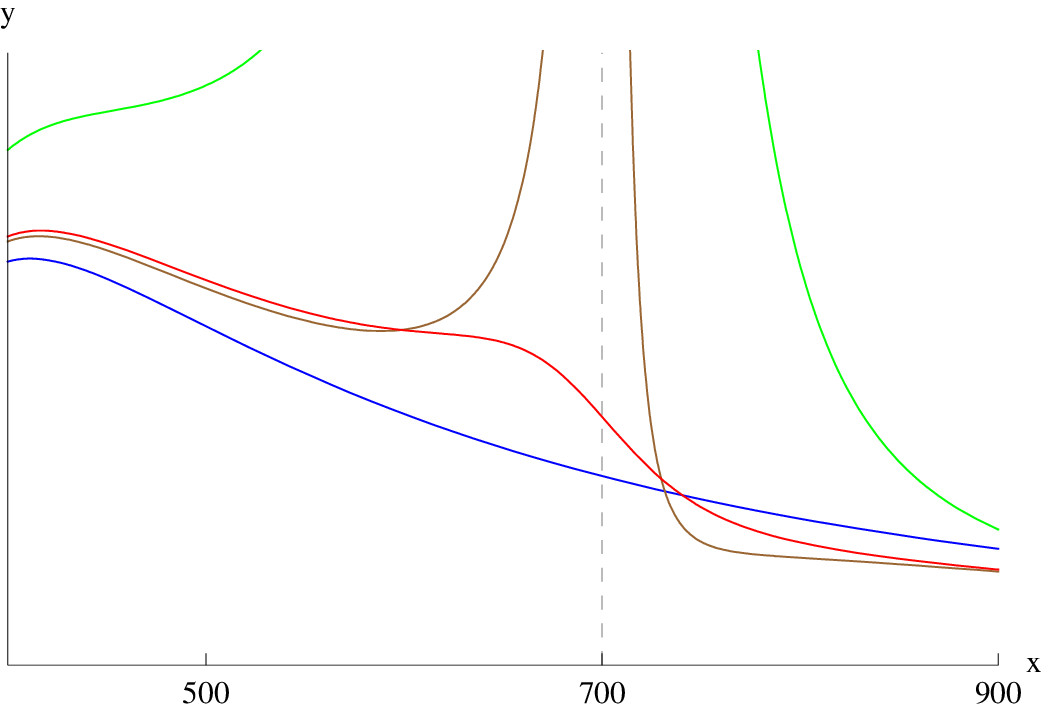}
\newline
(a)
\end{center}
\end{minipage}
\hspace{1.3cm}
\begin{minipage}[b]{0.44\linewidth}
\psfrag{y}{\scriptsize $\frac{dA^t_{FB}}{dM_{t \bar t}}$}
\begin{center}
\includegraphics[width=1\textwidth]{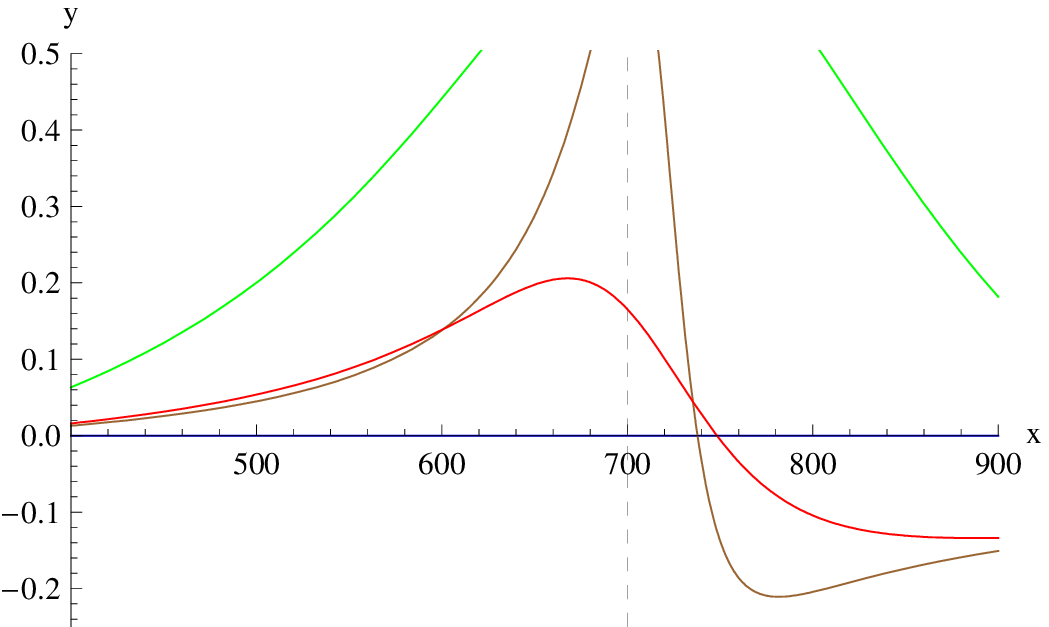}
\newline
(b)
\end{center}
\end{minipage}
\begin{center}
\begin{tabular}{|c|c|c|c|}
\hline
Colour & $f_{q_R}$ & $f_{t_R}$ & $\Gamma_{G^*}$ [GeV] \\
\hline
 ~Blue (SM)~ & ~0~ & ~0~ & ~-~ \\
 ~Green~ & ~-1~ & ~1~ & ~38~ \\
 ~Brown~ & ~-0.2~ & ~1~ & ~13~ \\
 ~Red~ & ~-0.05~ & ~5~ & ~143~ \\ 
\hline
\end{tabular}
\end{center}
\caption{Differential cross section in arbitrary units (a) and forward-backward asymmetry (b) at tree-level, as a function of the $t \bar t$ invariant mass for different light-quark ($f_{q_R}$) and top ($f_{t_R}$) Right couplings. (All Left couplings have been set
to zero and $f_{b_R}=1$.) The dashed lines correspond to the value of the $G^*$ mass ($M = 700$ GeV). From top to bottom at the left of both plots the lines colours are green, red, brown and blue. } 
\label{Figsa}
\end{center}
\end{figure}

The previous discussion is quite general for a gluon resonance with $M \sim 1$ TeV and $\Gamma_{G^*} / M \sim$ 0.2-0.3. Now, with the aim to illustrate the way we use that argument as a guide to generate the simulations presented in the next section, we take  a gluon resonance with $M = 700$ GeV as a typical example and analyze the role played by the couplings in the search of suitable $A^{t}_{FB}$ and $\sigma_{t \bar t}$ distributions. We also set, in this example, the quarks Left couplings to zero to simplify the analysis.    
In Fig.~\ref{Figsa}, we show the analytical predictions for $\sigma_{t\bar t}$ and $A_{FB}^t$ at the parton-level (these results have to be convoluted with the PDF's to obtain the physical values). For zero Right couplings, we obtain the SM cross section at tree-level (blue line) and a null contribution to the asymmetry. Now, to achieve the condition $a_q a_t < 0$, we need $f_{q_R}$ and $f_{t_R}$ to have  opposite signs --the other constraint, $v_q a_q v_t a_t > 0$, is then directly fulfilled--. Besides, as a relative large asymmetry is required, we set first both couplings to have an absolute value equal to 1. This selection results in too large deviations from measurements and it is forbidden (green lines). We reduce then the value of the light quark right coupling, $f_{q_R}=-0.2$, to produce a lower impact on the observables, however, this choice leads to a rather narrow resonance that still gives a large deviation in the resonant region (brown lines). Finally, in red lines, we find a possible solution. We see that the inadequate effect produced by a narrow-width may be changed by making the resonance broader. We keep a small value for $f_{q_R}$ but increase $f_{t_R}$ up to 5, giving a width $\Gamma \sim 150$ GeV, which is of the same order as the typical values for allowed parameter space points found in the simulations. Therefore, this analysis shows us the main ingredients to be taken into account and anticipates light-quark (top) chiral couplings of order $\sim 0.1\ (\gtrsim1)$. Besides, there could be new channels open, for example if new coloured fermions $Q$ were present, the processes $G^*\to q\bar Q$ and $G^*\to Q\bar Q$ could be open whenever the available phase space allows them~\cite{Barcelo:2011vk}. An alternative mechanism to explain the experimental data may occur in that case because $\Gamma_{G^*}$ would increase without the need of enlarging the couplings. In this work we assume that these channels are not open.

For large $G^*$ masses ($M \sim 2.5$ TeV), the INT term dominates in the whole $s$ range, always giving positive contributions to $A^{t}_{FB}$ and $\sigma_{t \bar t}$. In order to have an agreement with $A^{t}_{FB}$, the couplings of the light quarks have to be $\sim$ 1 because of the lack of a significant NPS contribution in the resonant region. Finally, the dependence on $\Gamma_{G^*}$ is smoother for heavy gluons than in the case of light gluons.

We see from Eqs.~(\ref{aes2}) and (\ref{aes3}) that the strength of the INT and NPS terms are controlled by the product of the light and top couplings. On the other hand, the $G^*$ width is given by a sum of light and top couplings squared (and a secondary dependence on the mass of the quarks). Therefore, when exploring the parameter space we have that if --as it is going to be the case-- the light couplings are small and the top couplings large, then modifications to the width arise mainly from changes in the top couplings. On the other hand, variations in the strength of INT and NPS terms are equally sensitive to changes in both the light and top couplings. Therefore, in a first approximation, if we need to modify the strength of the INT and NPS contributions without producing large variations in the $G^*$ width, we should only change the light-quark couplings; if we need to modify the width, we should change the top couplings. This discussion will be helpful for the analysis performed in the next section.


\section{Numerical scan and results} 
\label{numeric}

In this section we perform a numerical scan of the regions of parameter space pointed out by the previous paragraphs analytic study of the $t\bar t$ forward-backward asymmetry ($A_{FB}^t$) and invariant-mass spectrum in the $q\bar q \to t \bar t$ process.  In this numerical scan we use {\tt Madgraph/Madevent} \cite{mgme} (MGME)~\footnote{We have slightly modified {\tt MGME} to include a running width for the resonance.} with CTEQ6L parton distribution functions~\cite{Pumplin:2002vw} to simulate Tevatron $p\bar p \to t\bar t$ process and compute $A_{FB}^t$ and \mtt  at parton-level. 

In order to test the different points in the parameter space, we compare our simulated $A_{FB}^t$ results to the CDF two-bins unfolded result on this observable \cite{CDF3-Asym}.  On the other hand, given the lack of unfolded results in the recent experimental data on the \mtt \cite{CDF3-Asym}, we interpret the experimental data as that there is agreement with the SM expectation. Therefore, we simulate the \mtt using our model and using the SM and compare both results. In this last comparison we use in both simulations the same values for the top mass, and the factorization and renormalization scale.  (In any case, we have checked that  changes in these values within their allowed limits do not modify our results within the statistical error.) 

In order to compare the forward-backward asymmetry results, we perform a $\chi^2$ test on the two bins of the unfolded data for $A_{FB}^t$.  To compare the $t\bar t$ invariant-mass spectrum we also perform a $\chi^2$ test, but as follows. We divide the $M_{t\bar t}$ domain in $15$ GeV bins up to $M_{t\bar t}=700$ GeV and from there up we divide in three bins in $740$, $800$ GeV and greater than $800$ GeV, in order to assure at least 5 events per bin in the SM simulated \mtt with $5.3\ \mbox{fb}^{-1}$ (the collected luminosity in the CDF results).  To avoid large fluctuations we always simulate $100.000$ events, and for the \mtt we scale the results to the measured luminosity.  In each $\chi^2$ test we require $p>0.05$ to state agreement of the NP model on the corresponding observable.

We perform a scanning of two different regions in parameter space.  In the first case, in order to obtain clean results, we set all the Left couplings of the quarks to the new gluon to zero, and plot our results solely as a function of the light and top quarks Right couplings.  The motivation for this choice in the top quarks is simplicity and the existence of precision measurements in $B$-physics \cite{b-precision} which constrain the couplings of the $(t_L\ b_L)$ doublet to new gluons.  In this scenario we have that axial and vector couplings are equal; therefore the asymmetry at the resonance is positive (see Eq.~\ref{aes3}).  By setting opposite signs to the light and top Right couplings we get a positive pre-peak in the asymmetry, as sought.  In the second case we perform a random scan of all couplings in order to see the regions which have better agreement with $A_{FB}^t$ and the \mtt\!\!.

\begin{figure}[!htbp]
\begin{center}
\psfrag{qR}{$f_{q_R}$}
\psfrag{tR}{$f_{t_R}$}
\begin{minipage}[b]{0.45\linewidth}
\begin{center}
\includegraphics[width=1\textwidth]{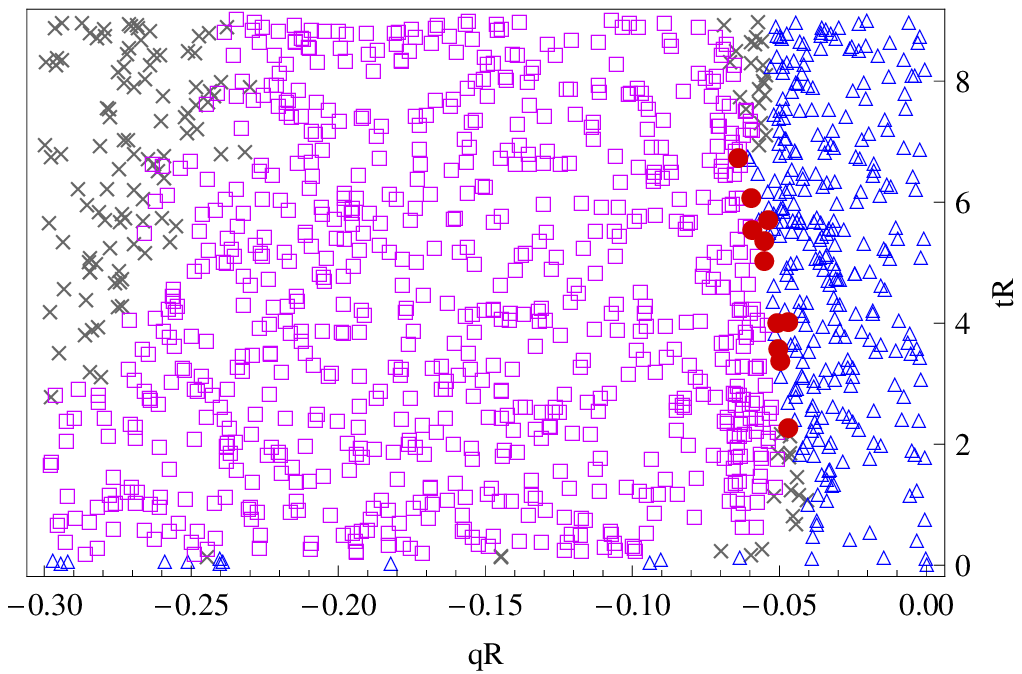}
\newline
(a) $M=700$ GeV
\end{center}
\end{minipage}
\hspace{0.5cm}
\begin{minipage}[b]{0.45\linewidth}
\begin{center}
\includegraphics[width=1\textwidth]{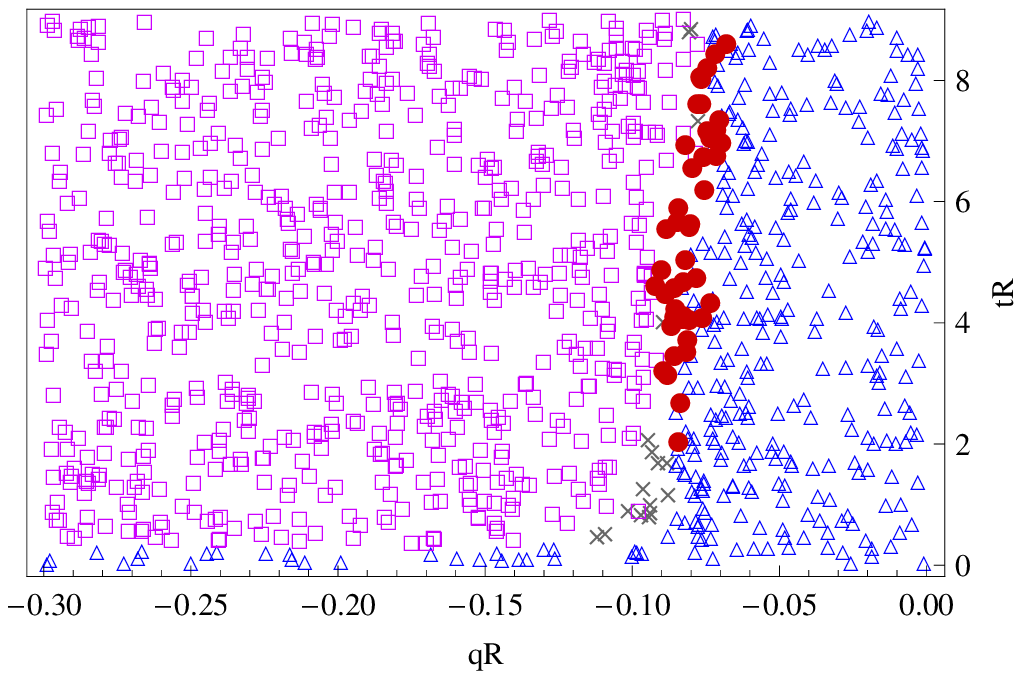}
\newline
(b) $M=850$ GeV
\end{center}
\end{minipage}
\begin{minipage}[b]{0.45\linewidth}
\begin{center}
\includegraphics[width=1\textwidth]{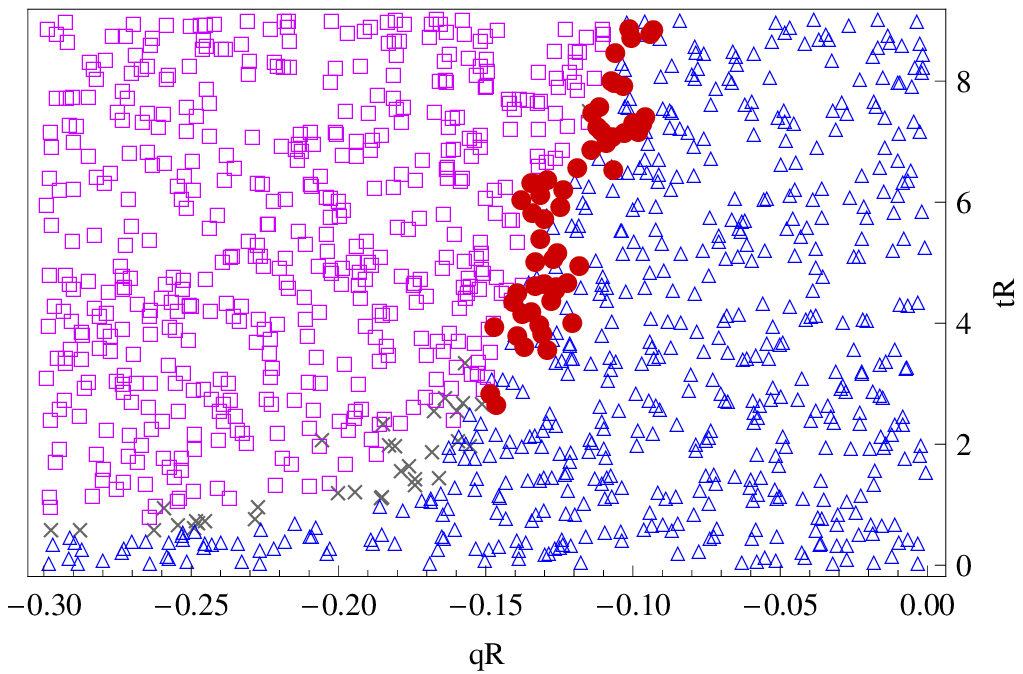}
\newline
(c) $M=1000$ GeV
\end{center}
\end{minipage}
\hspace{0.5cm}
\begin{minipage}[b]{0.45\linewidth}
\begin{center}
\includegraphics[width=1\textwidth]{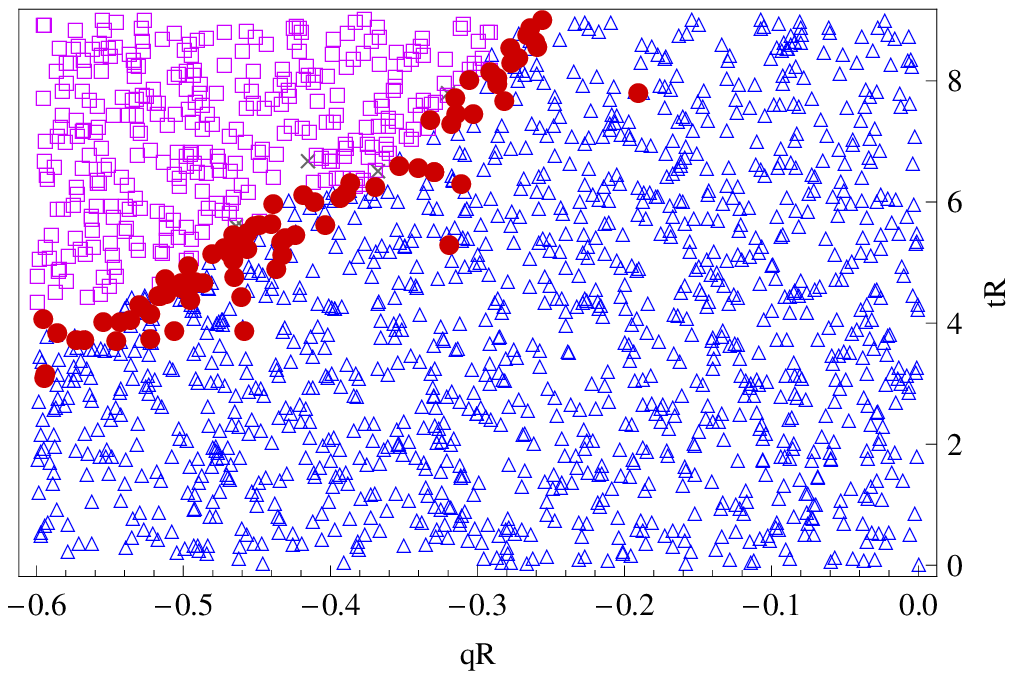}
\newline
(d) $M=1500$ GeV
\end{center}
\end{minipage}
\caption{Results of the numerical scan where we test agreement with $A_{FB}^t$ and the $M_{t\bar t}$-spectrum at the $p\lessgtr 0.05$ level for different masses of the gluon partner. $f_{q_R}$ stands for all the light quarks ($u,\ d,\ c,\ s$ and $b$) Right couplings, and $f_{t_R}$ for the top Right coupling (all in units of $g_{s}$); all Left couplings are set to zero in this scan. Black crosses correspond to no agreement at all, magenta squares to only agreement in $A_{FB}^t$, blue triangles to only agreement in the $M_{t\bar t}$-spectrum and red points to agreement in both $A_{FB}^t$ and the $M_{t\bar t}$-spectrum.  In all plots the random density is uniform everywhere, but in the red-points region, where it has been doubled in order to obtain better contrast. Notice the different scale in plot (d).} 
\label{leftzero}
\end{center}
\end{figure}

In Figure \ref{leftzero} we plot the results for the scan where all the Left couplings are set to zero.  We have performed a uniform density random scan in this region and use the black crosses to state no agreement neither in $A_{FB}^t$ nor in \mtt\!\!, the magenta squares to state only $A_{FB}^t$ agreement, the blue triangles to state only \mtt agreement and the red points to state agreement in both observables.  After the first scan we have doubled the random density close to the red points region in order to increase the contrast therein.

These plots should be read as follows. First, observe that the SM, which is the $f_{q_R}=f_{t_R}=0$ point at the origin, does not agree with the forward-backward asymmetry. Then notice that as couplings increase (in absolute value) from zero, the only agreement is, as expected, with \mtt (blue triangles).  For larger couplings this agreement has to break down.  On the other hand, from the top and left region of the plots we have --as expected from previous section-- agreement with the forward-backward asymmetry (magenta squares). (In Figure \ref{leftzero}a the top left corner begins with no agreement, since such a light resonance gives too large asymmetry for those couplings.)  Again, this agreement has to break down as couplings decrease in absolute value.  Henceforth, these two regions (blue triangles and magenta squares) could be joined mainly in two ways.  Either through a region where both agreements ($A_{FB}^t$ and $M_{t\bar t}$-spectrum) fail (black crosses), or either they overlap and there is an agreement in both observables (red points).

We see from the plots in Fig.~\ref{leftzero} that, as it was the first motivation of the work, there is a strip of favourable (red) points for a light resonance beginning at $M\gtrsim 700$ GeV.  Although in all cases the agreement region is a relatively thin strip, this agreement is slightly more convincing for larger $M$.  It is interesting to observe at this point that the favourable region that has raised from the phenomenological analysis is composite models friendly, with only the Right chirality of the top quark being composite in this case, since it has large couplings to New Physics. A possible way to broaden the red strip is to consider the presence of extra fermions $Q$ that open new decay channels modifying the resonance width \cite{Barcelo:2011vk}. However,
for the case of light $Q$ fermions, a deeper examination of the existing collider phenomenology should be carried on.

\begin{figure}[!htbp]
\psfrag{qR}{$f_{q_R}$}
\psfrag{tR}{$f_{t_R}$}
\psfrag{qL}{$f_{q_L}$}
\psfrag{tL}{$f_{t_L}$}
\begin{center}
\begin{minipage}[b]{0.35\linewidth}
\begin{center}
\includegraphics[width=1\textwidth]{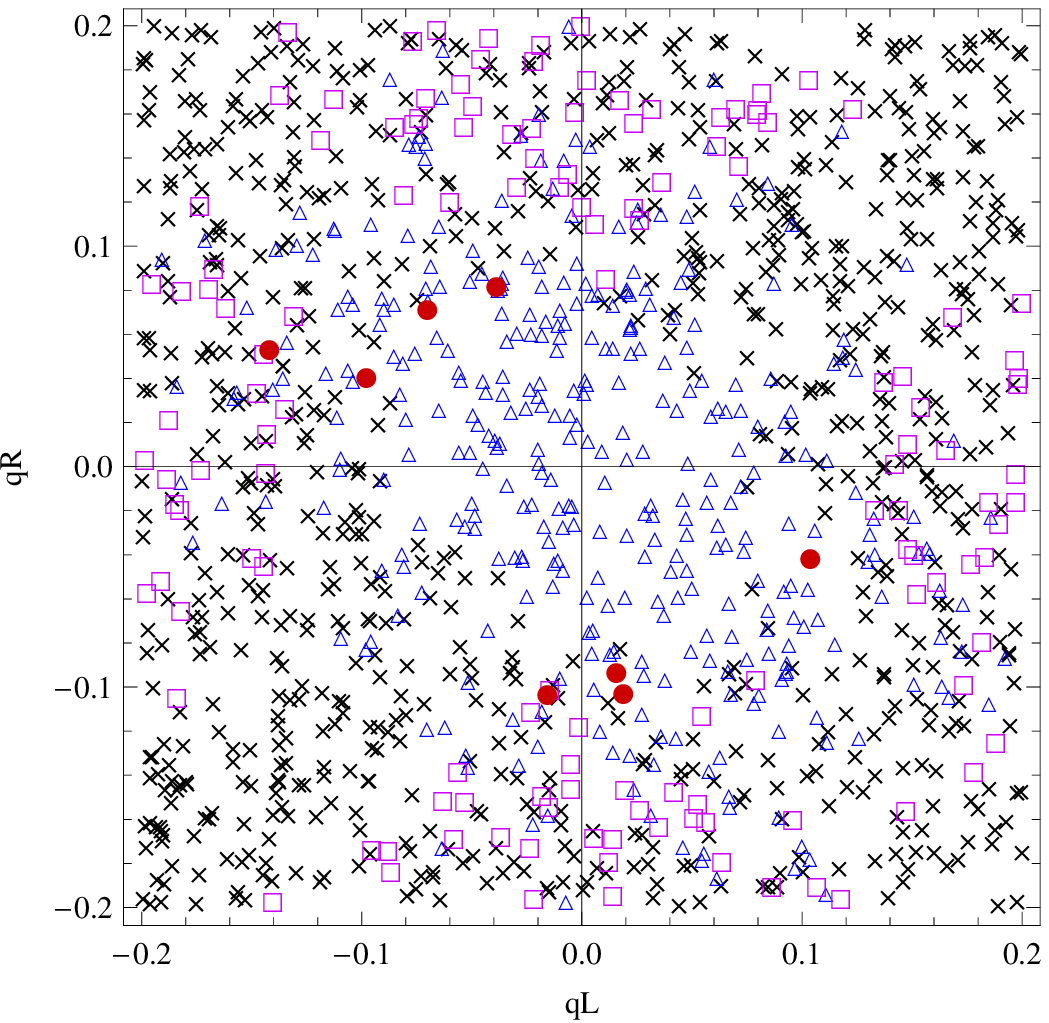}
\newline
(a) $M=700$ GeV
\end{center}
\end{minipage}
\hspace{1.5cm}
\begin{minipage}[b]{0.35\linewidth}
\begin{center}
\includegraphics[width=1\textwidth]{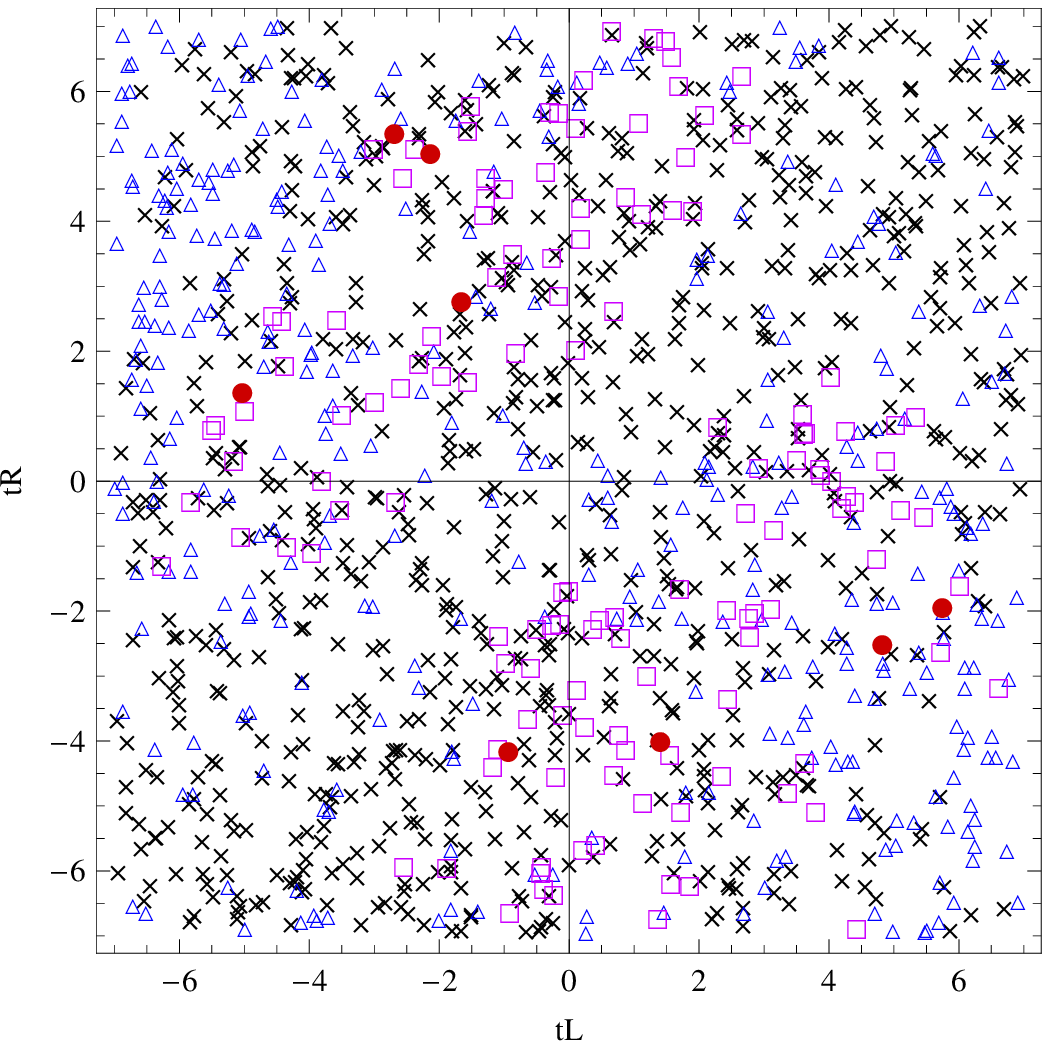}
\newline
(b) $M=700$ GeV
\end{center}
\end{minipage}
\begin{minipage}[b]{0.35\linewidth}
\begin{center}
\includegraphics[width=1\textwidth]{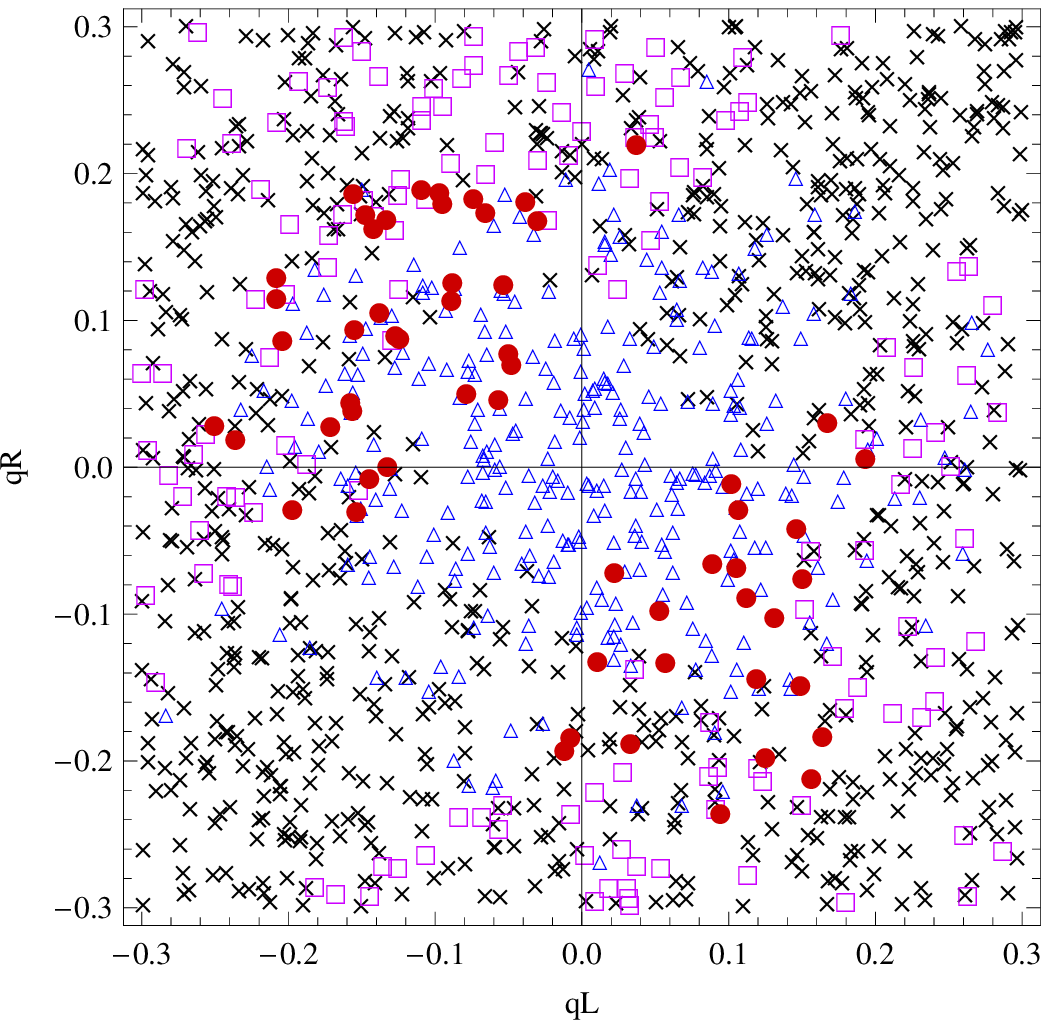}
\newline
(c) $M=850$ GeV
\end{center}
\end{minipage}
\hspace{1.5cm}
\begin{minipage}[b]{0.35\linewidth}
\begin{center}
\includegraphics[width=1\textwidth]{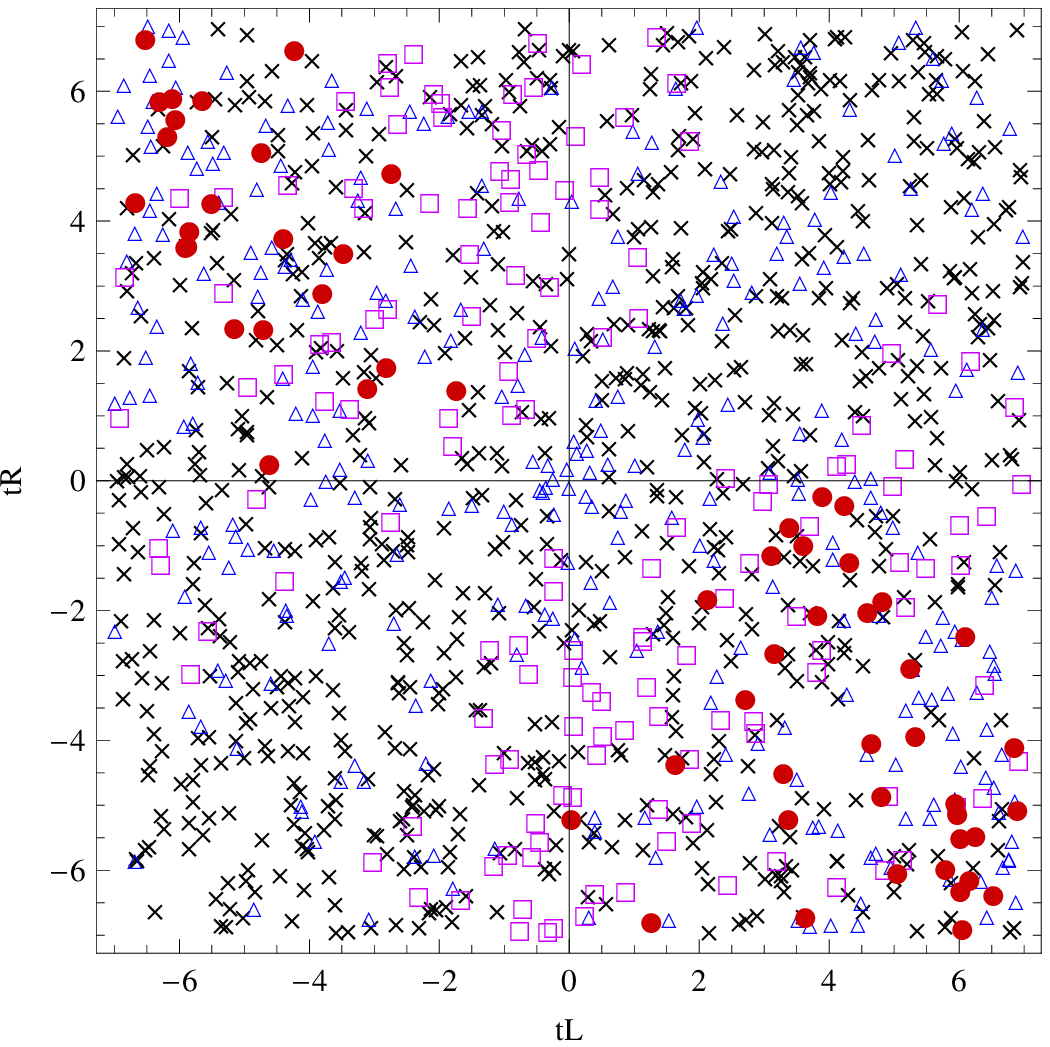}
\newline
(d) $M=850$ GeV
\end{center}
\end{minipage}
\begin{minipage}[b]{0.35\linewidth}
\begin{center}
\includegraphics[width=1\textwidth]{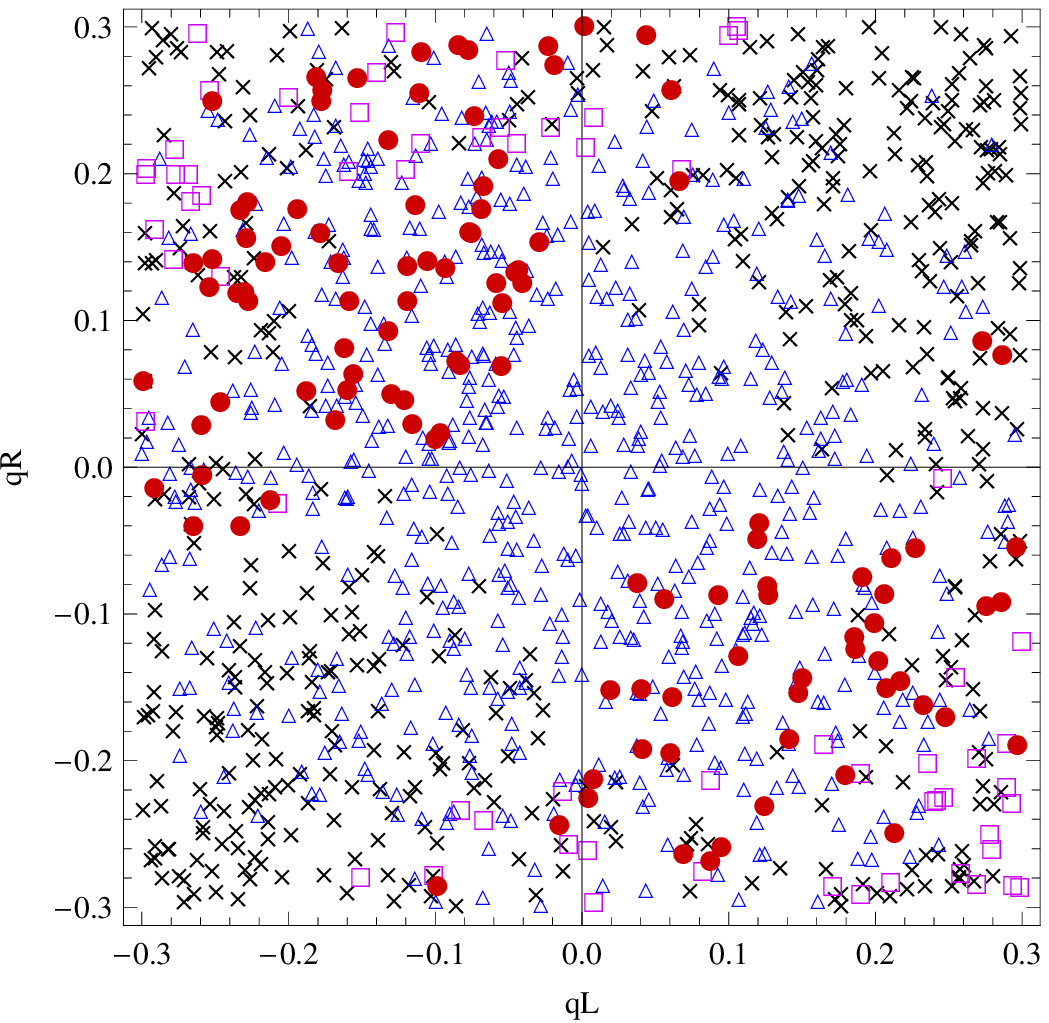}
\newline
(e) $M=1000$ GeV
\end{center}
\end{minipage}
\hspace{1.5cm}
\begin{minipage}[b]{0.35\linewidth}
\begin{center}
\includegraphics[width=1\textwidth]{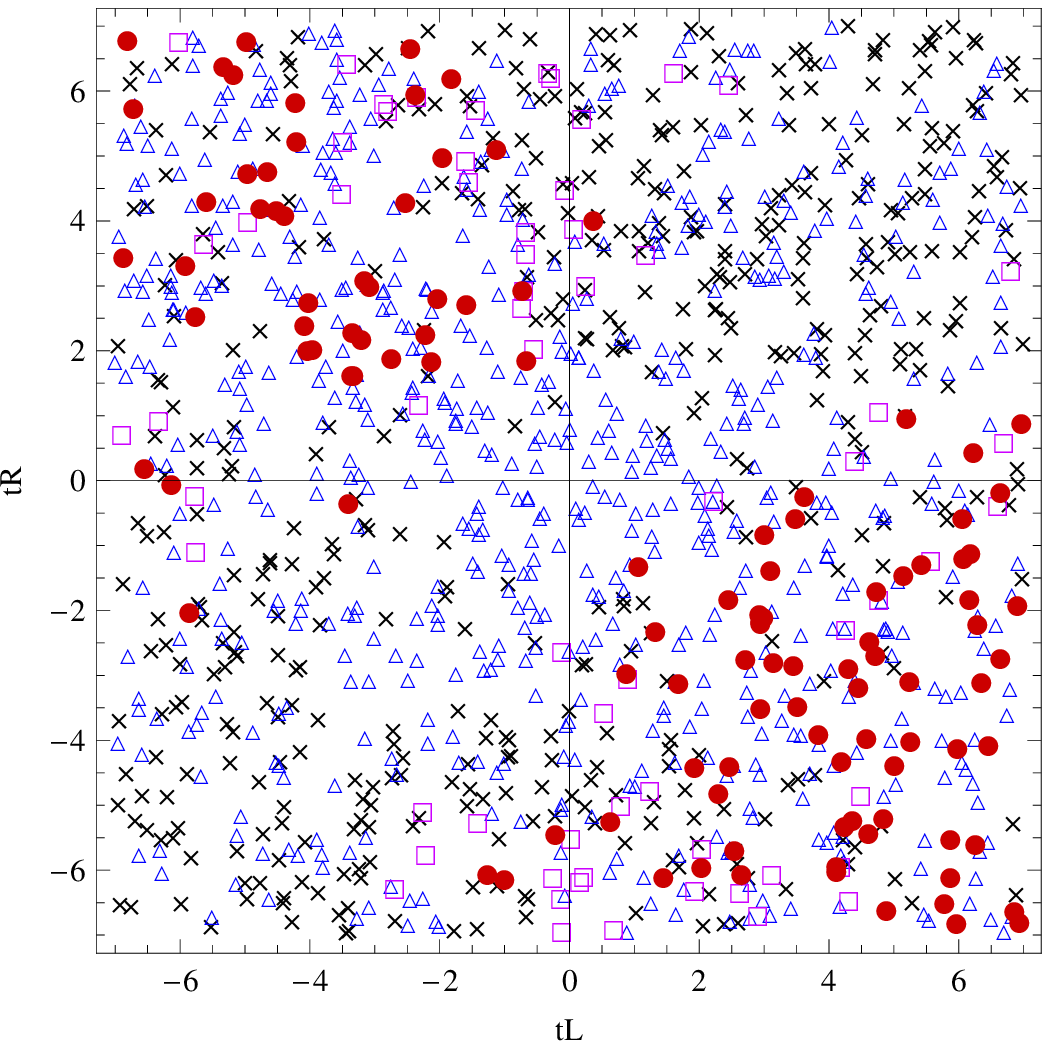}
\newline
(f) $M=1000$ GeV
\end{center}
\end{minipage}
\caption{Exploration of the parameter space through a total random scan.  $f_{b_R}$ (which is not plotted) takes random values between $-2$ and $+2$, and all the couplings are expressed in units of $g_{s}$.  In each plot there are three variables which are not plotted and produce a mixture between the different colour regions. Black crosses correspond to no agreement at all, magenta squares to only agreement in $A_{FB}^t$, blue triangles to only agreement in $M_{t\bar t}$-spectrum and red points to agreement in both $A_{FB}^t$ and $M_{t\bar t}$-spectrum.} 
\label{random}
\end{center}
\end{figure}

At this point, it is also interesting to understand the red points
patterns in Fig.~\ref{leftzero} from the previous section parton-level
equations (\ref{aes2}) and (\ref{aes3}).  The off-resonant case (large
$M$) is easy to understand, since in this case the dominant contribution to
$t\bar t$-production comes from the INT term, Eq.~(\ref{aes2}).  As
previously discussed, the effect of the width in this term may be
neglected in a first approximation and, therefore, the dependence on
the couplings is just through the $f_{q_R} f_{t_R}$ product.
Therefore, regions where this product remains constant will have
approximately the same outcome for $A_{FB}$ and \mtt.  In
fact, this is easily verified for $M=1500$ GeV in
Fig.~\ref{leftzero}d, where the red points strip is along a $f_{q_R}
f_{t_R} \simeq $ constant line.

On the other hand, in the resonant case (low $M$) it is the NPS term,
Eq.~\ref{aes3}, the one that rules, at least close to the resonance.
In this case the width plays a major role and it has to be taken into
account.  We observe from Eq.~\ref{aes3} that, for the case of large
$f_{t_R}$ and little $f_{q_R}$, the NPS term close to the resonance
goes essentially as $f_{q_R}^2 f_{t_R}^2/f_{t_R}^4$.  Therefore, an
increase in $|f_{t_R}|$ will have the effect of diminish and broaden
the peak in the \mtt and $A_{FB}$.  Since the
resolution in $A_{FB}$ is only in two large bins, then the change will
be minor here.  Henceforth, we may expect that if only $|f_{t_R}|$ is
increased from a favourable (red) point in parameter space, then the
agreement with \mtt and $A_{FB}$ may hold.  On the other
hand, if we only move $f_{q_R}$ we will have that the width remains
unchanged and the NPS changes to disfavourable regions.  In fact, if
$|f_{q_R}|$ is decreased, then the peak in $A_{FB}$ decreases but is
not broaden, therefore spoiling the asymmetry.  If $|f_{q_R}|$ is
increased then the peak in \mtt is also increased and spoils
this observable.  These observations are best seen in
Figs.~\ref{leftzero}a and \ref{leftzero}b for $M=700$ and $850$ GeV,
where the red points lie along a vertical strip.

\begin{figure}[!htbp]
\psfrag{qR}{$f_{q_R}$}
\psfrag{tR}{$f_{t_R}$}
\psfrag{qL}{$f_{q_L}$}
\psfrag{tL}{$f_{t_L}$}
\begin{center}
\begin{minipage}[b]{0.35\linewidth}
\begin{center}
\includegraphics[width=1\textwidth]{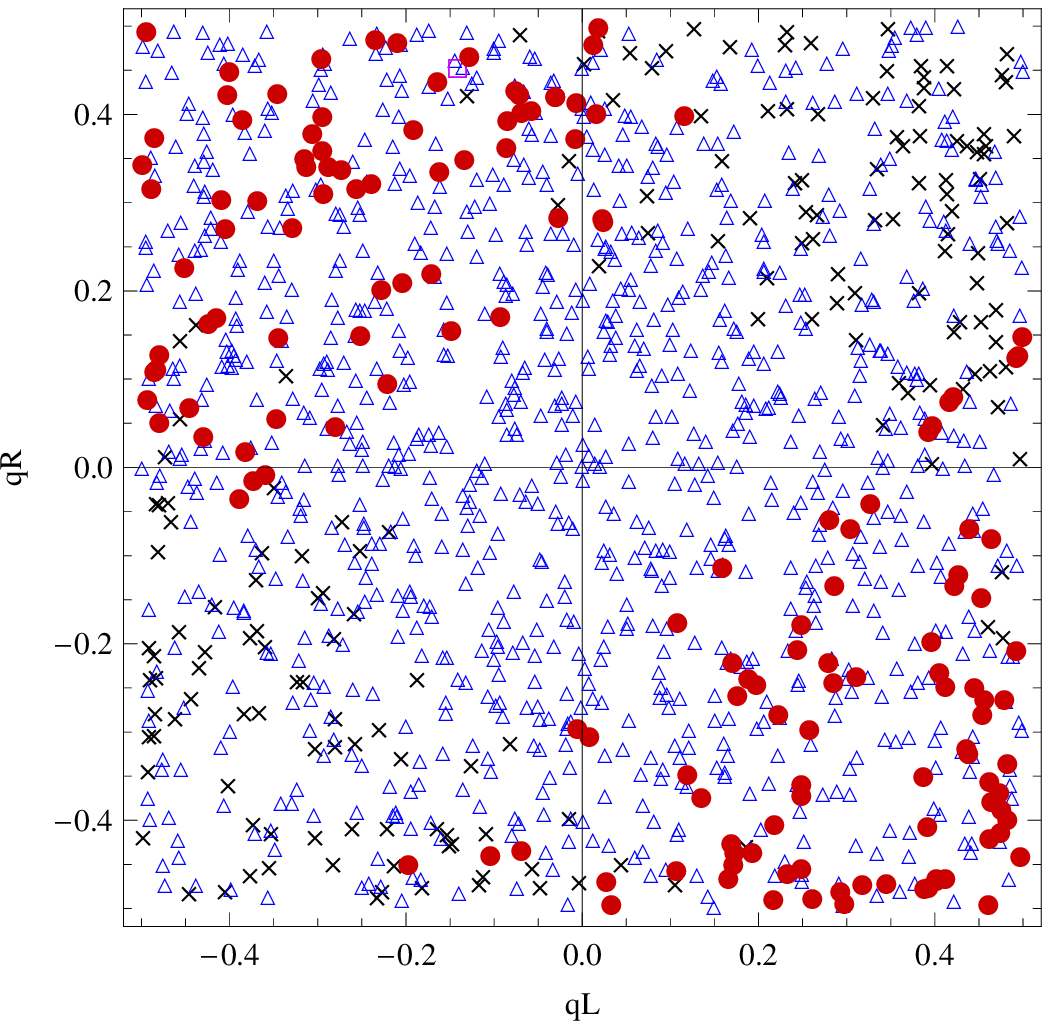}
\newline
(a) $M=1500$ GeV
\end{center}
\end{minipage}
\hspace{0.5cm}
\begin{minipage}[b]{0.35\linewidth}
\begin{center}
\includegraphics[width=1\textwidth]{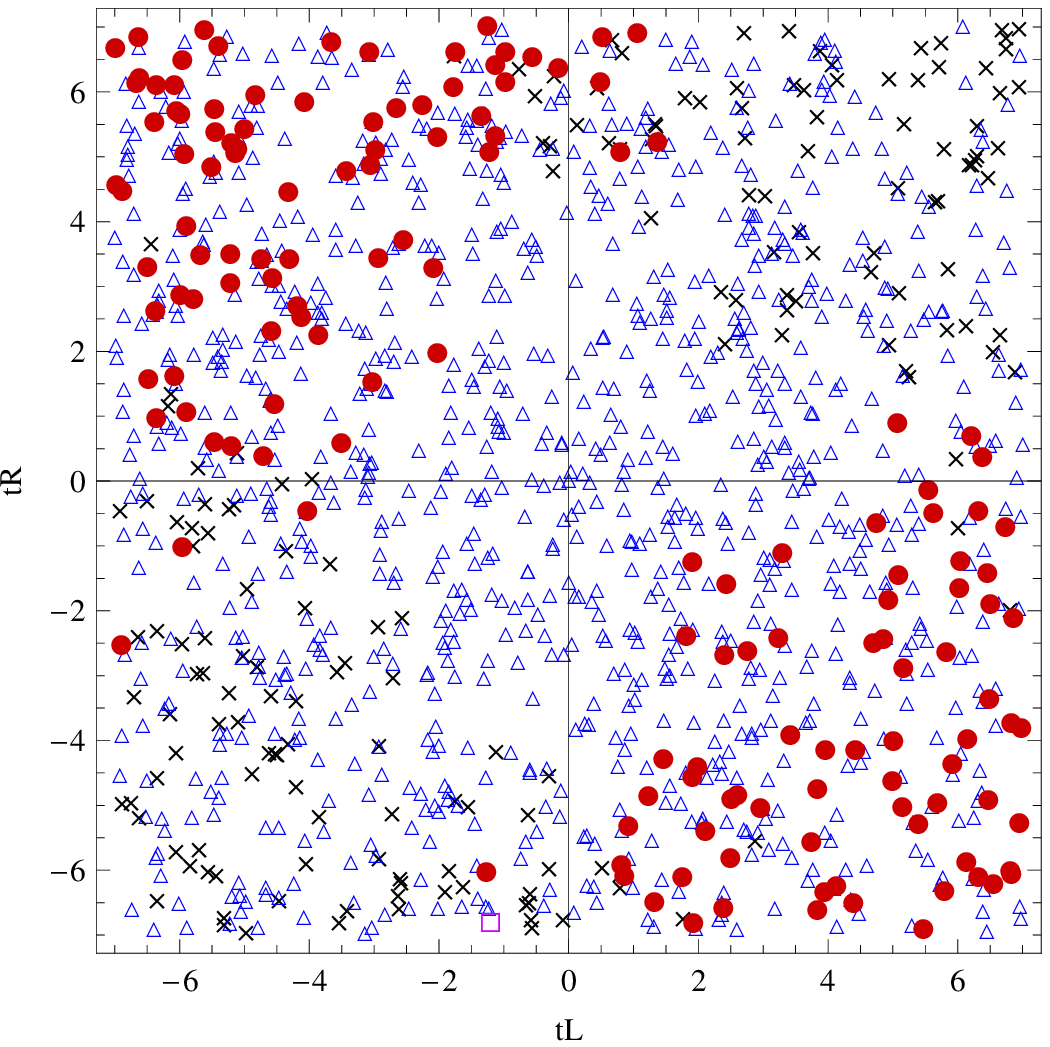}
\newline
(b) $M=1500$ GeV
\end{center}
\end{minipage}
\begin{minipage}[b]{0.35\linewidth}
\begin{center}
\includegraphics[width=1\textwidth]{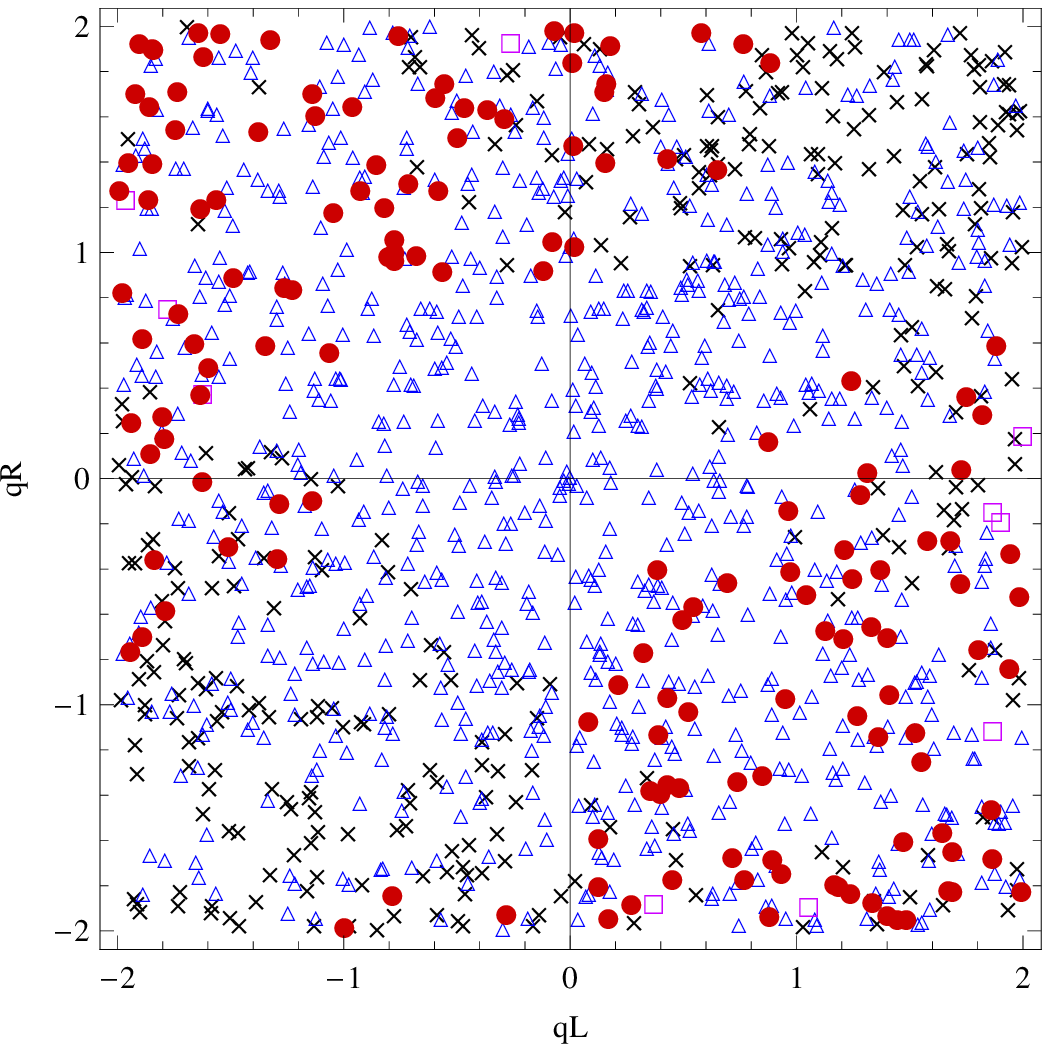}
\newline
(c) $M=2500$ GeV
\end{center}
\end{minipage}
\hspace{0.5cm}
\begin{minipage}[b]{0.35\linewidth}
\begin{center}
\includegraphics[width=1\textwidth]{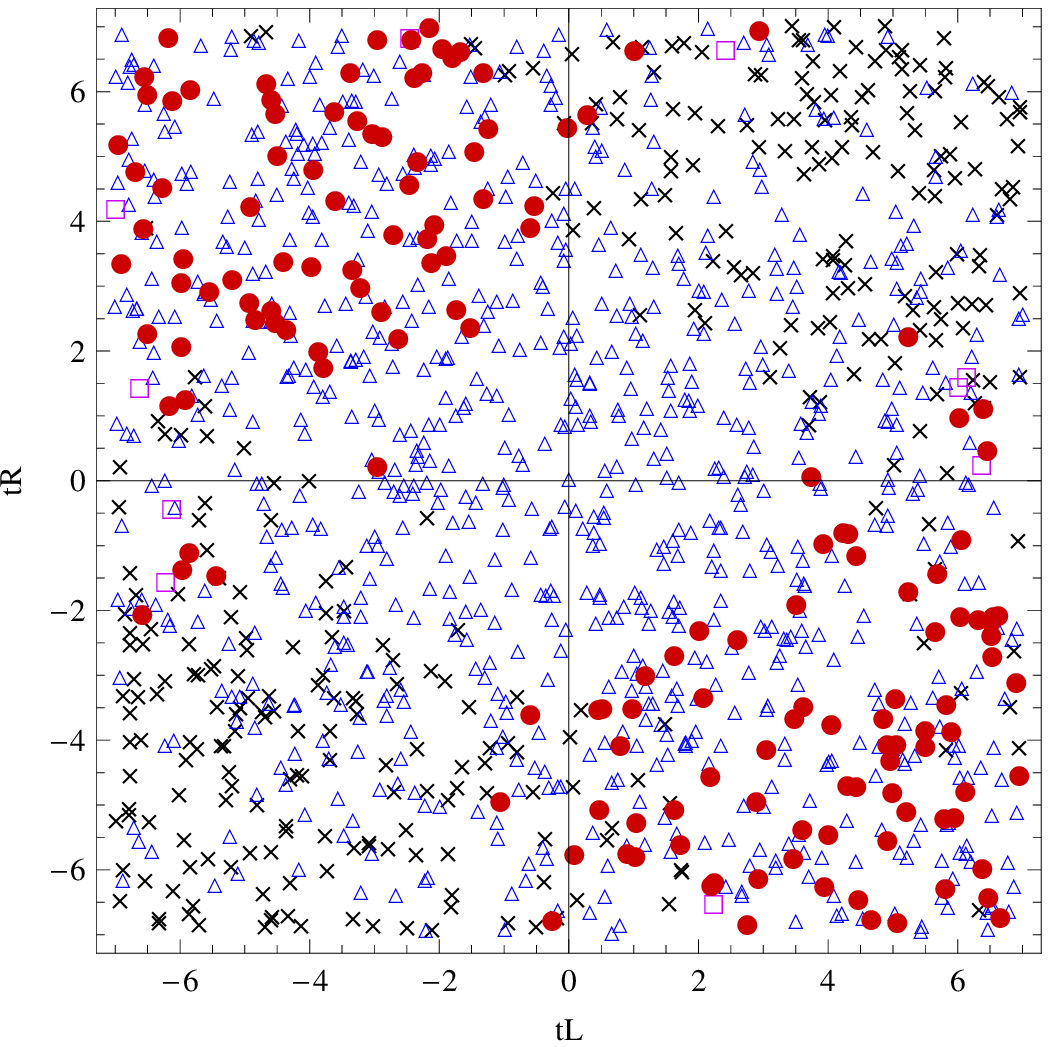}
\newline
(d) $M=2500$ GeV
\end{center}
\end{minipage}
\caption{Same analysis as in Fig.~\ref{random}, but for heavier masses which could not produce a resonance at Tevatron. Black crosses correspond to no agreement at all, magenta squares to only agreement in $A_{FB}^t$, blue triangles to only agreement in $M_{t\bar t}$-spectrum and red points to agreement in both $A_{FB}^t$ and $M_{t\bar t}$-spectrum.}
\label{random2}
\end{center}
\end{figure}

Finally, we have also performed a total random scan of the couplings.  We present the results of this scan for masses of the resonance which could be resonant at Tevatron ($M=700,\ 850$ and $1000$ GeV) in Fig.~\ref{random}, and for larger masses ($M=1500$ and $2500$ GeV) in Fig.~\ref{random2}.  In all cases we have varied $f_{q_L}$, $f_{q_R}$, $f_{t_L}=f_{b_L}$, $f_{t_R}$ and $f_{b_R}$.  The variation of $f_{b_R}$, which is between $-2$ and $+2$ (in units of $g_{s}$) and is not plotted, modifies slightly the width of $G^*$ and henceforth the outcome of the simulation.  At difference with the previous case, in each of these plots there are three couplings which change randomly and are not being plotted.  Therefore, there is not a clear border between the different colour regions in this case.  A general feature of all plots is that the favourable points show an axial-like preference for the couplings of light and top quarks. Moreover, we found that, as anticipated in the previous section, these axial couplings have in general opposite signs (this result cannot be seen from the plots).  Also notice that, as expected, as $M$ increases also the couplings required to satisfy the CDF observables increase (in absolute value).  And last, observe that the density of red points increases considerably as $M$ increases, which would mean that a large $M$ solution is favoured.

As predicted in the previous section, as the mass of $G^*$ increases, the NPS term becomes less important and therefore the large-width requirement to preserve the \mtt agreement may be relaxed. This is translated in Figs.~\ref{random} and \ref{random2} in that magenta squares density decreases (where $A_{FB}^t$ agrees, but \mtt fails) and red points density increases ($A_{FB}^t$ and \mtt agree), as $M$ increases from $700$ GeV to $2500$ GeV.

\section{LHC phenomenology}
\label{lhc}

Our analysis shows that a light gluon resonance with $M\gtrsim700$ GeV
can accommodate the CDF
data on $t\bar t$ production within $p>0.05$. However, there are new
data from LHC that show agreement with the SM predictions. The most
important constraints for this type of gluon resonance searches arise
from the $t\bar t$ production \cite{ttbarAtlas35,ttbarAtlas200} and
dijet final states \cite{Aad:2011aj,updatedijet} in LHC at 7 TeV. In
this section we analyze those constraints for all the points of the
parameter space that reproduce the Tevatron measurements, using the
random scan of section~\ref{numeric}. We also show the main
predictions of our model for LHC.

Concerning top-pair production, Ref.~\cite{ttbarAtlas35} shows
agreement of the cross section with the SM for an accumulated
luminosity of 35 pb$^{-1}$, whereas the recent analysis of
\cite{ttbarAtlas200} sets limits on the production of resonances with
an accumulated luminosity of 200 pb$^{-1}$. Ref.~\cite{ttbarAtlas200}
excludes a gluon resonance with $M\lesssim 650$ GeV for a specific
choice of couplings: $f_{q_L}=f_{q_R}=f_{b_R}=-0.2$,
$f_{t_L}=f_{b_L}=1$ and $f_{t_R}=4$. Note that, as shown in the
previous sections, the Tevatron data disfavours vector-like couplings
as the ones chosen for the light quarks in \cite{ttbarAtlas200}, since
they lead to too low $A_{FB}^t$. Since the Tevatron $A_{FB}^t$
requires large couplings between the top-quark and the resonance,
$f_t\gtrsim 1$, the top-pair pair production is one of the better
tests for a gluon resonance of this kind at LHC. We have simulated the
production cross section of a gluon resonance decaying to $t\bar t$ at
LO at parton-level, finding $\sigma\lesssim 2.5$ pb in all the cases.
Since the analysis of \cite{ttbarAtlas200} gives upper bounds for
$M\lesssim 1.6$ TeV only, we can not precisely check limits in the
case of $M=2.5$ TeV. For a light resonance mass, $M\lesssim 1$ TeV,
the limits set by \cite{ttbarAtlas200} are well beyond the results of
our model, mainly because the Tevatron results can be obtained with
small couplings for the light quarks, suppressing the $t\bar t$
production through $G^*$ at LHC. For $M\sim 1.5$ TeV, the ATLAS
sensitivity reported in \cite{ttbarAtlas200} is near the region of the
parameter space that can explain $A_{FB}^t$. We expect that with an
accumulated luminosity greater than 1 fb$^{-1}$, either an excess in
$\sigma_{t\bar t}$ can be measured or a large region of the parameter
space with $M\sim 1.5-2.5$ TeV can be excluded, mostly the region with
larger top-couplings. Also a better resolution of the invariant mass
distribution could be useful to test our predictions for $t\bar t$.

In Ref.~\cite{Aad:2011aj} the ATLAS collaboration has provided
experimental results on the search for new resonances in dijet final
states with 36 pb$^{-1}$. A recent update with 163 pb$^{-1}$ provides
stronger constraints on the search for new resonances and extends the
analysis to larger masses~\cite{updatedijet}. A light gluon resonance
can give large contributions to jet pair creation, as shown in Fig.~2
of Ref.~\cite{updatedijet} for axial couplings of order 1. Since the
Tevatron data selects the couplings of the light quarks, we analyze
the predictions for dijet production at LHC, that have a strong
dependence with the resonance mass. First notice that, since the events analyzed by ATLAS include $b$-jets, the dijet
production through $G^*$ is controlled by two terms, one including only light-jets proportional to $f_q^4$ and another including $b$-jets proportional to $f_q^2\times f_b^2$. For $M=\{700,850,1000,1500\}$ GeV, we find agreement with the Tevatron
data for regions of the parameter space with $f_q\lesssim\{0.1,0.2,0.3,0.4\}$, whereas $f_b\sim 1$. These couplings lead to a suppression factor $\sim10^{-4}-10^{-2}$ for light-jets, and a suppression factor $\sim10^{-2}-10^{-1}$ for $b$-jets, with a dispersion that depends on the precise value of $f_b$. On the other hand, for $M=2.5$ TeV, larger $f_q$ are required to reproduce the Tevatron results on $t\bar t$, $f_q\sim 0.5-2$, that is similar to the range for $f_{b_R}$, but smaller than the range for $f_{b_L}=f_{t_L}$. Therefore in this case, although there is a large suppression from $M$, the large couplings induce a sizable contribution that can be tested with the present data.
We have simulated at LO the dijet production in our model with {\tt MGME} and {\tt Pythia}~\cite{Sjostrand:2006za} for hadronization and showering simulations. We have considered the following set of kinematic cuts, implemented in~\cite{updatedijet}: $|\eta_j|<2.5$, $|\Delta\eta_{jj}|<1.3$, $p_T^{j1}>180$ GeV, $p_T^{j2}>30$ GeV and
$M_{jj}>700$ GeV. Although we have not made a detailed simulation of the detectors, neither computed NLO corrections, we expect that our LO predictions can give the correct order of magnitude, mainly on the upper bound limits. Our results show that, for $M\lesssim 1.5$ TeV, the dijet limits on the cross section are all beyond the present sensitivity.  For $M=2.5$ TeV the ATLAS analysis put severe constraints on the parameter space. Taking the $95\%$ confidence level upper limit
of ~\cite{updatedijet}, we obtain that a large region of the parameter space is ruled out in this case.

Let us comment also on the proportion of events with light- and $b$-jets. For $M=700$ and 850 GeV we find that the ratio of events with $b$-jets is between $95-99\%$ for the different points of the parameter space, whereas for $M=1000$ GeV the ratio of $b$-jets is $90-99\%$ and for $M=1500$ GeV the ratio is $40-99\%$. The dispersion in the frequency of $b$-jets is due to the different values of $f_q$ and $f_b$ for each point of the parameter space with fixed mass. On the other hand, for $M=2.5$ TeV, we find points of the parameter space where the ratio of $b$-jets is only $1\%$, and we also find points where this ratio raises to $99\%$, depending on the region of the parameter space. As expected, we also find that, for all the resonance masses, the points of the parameter space that have a lower ratio of events with $b$-jets correspond to lower values of $f_{b_L}$. For these reasons it will be very interesting to have experimental data on resonance searches with dijets and $b$-tagging, that could help to better select the favourable region of the parameter space.

\begin{figure}[ht]
\hspace*{-0.8cm}
\begin{minipage}[t]{0.68\linewidth}
\psfrag{-}[l]{\footnotesize }
\psfrag{3.50656}[l]{\hspace*{-0.2cm}\footnotesize 0.03}
\psfrag{2.29263}[l]{\hspace*{-0.2cm}\footnotesize 0.1}
\psfrag{1.20397}[l]{\hspace*{-0.2cm}\footnotesize 0.3}
\psfrag{0}[r]{\hspace*{-0.2cm}\footnotesize 1}
\psfrag{1.09861}[l]{\hspace*{-0.05cm} \footnotesize  3}
\psfrag{2.30259}[l]{\hspace*{-0.2cm} \footnotesize 10}
\psfrag{jj}[b]{\small$\sigma_{G^*}\times A_{jj}$\ \ [pb]}
\psfrag{tt}[t]{\small$\sigma_{G^*}\times BR(G^*\to t\bar t)$\ \ [pb]}
\psfrag{0.0}[t]{\footnotesize0}
\psfrag{0.5}[t]{\footnotesize0.5}
\psfrag{1.0}[t]{\footnotesize1}
\psfrag{1.5}[t]{\footnotesize1.5}
\psfrag{2.0}[t]{\footnotesize2}
\psfrag{2.5}[t]{\footnotesize2.5}
\psfrag{3.0}[t]{\footnotesize3}
\begin{center}
\includegraphics[width=.55\textwidth]{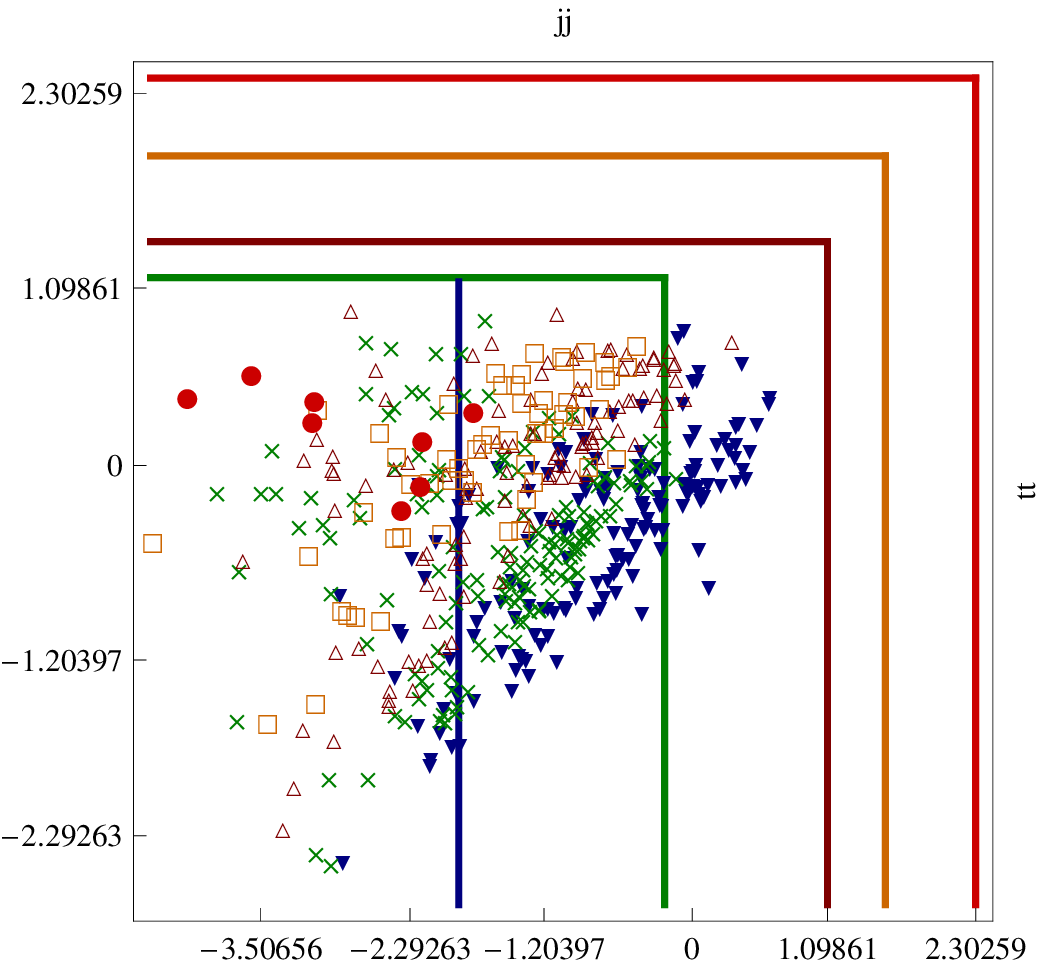}
\newline
\vskip0.01cm
\hspace*{-1.2cm} (a)
\end{center}
\end{minipage}
\hspace*{-3.5cm}
\begin{minipage}[t]{0.68\linewidth}
\psfrag{AC}[t]{\small$10\times A_C^{(NP)}$}
\psfrag{AFB}[b]{\small$A_{FB}^{t(NP)}$}
\psfrag{0.05}[tl]{\footnotesize0.05}
\psfrag{0.10}[tl]{\footnotesize0.10}
\psfrag{0.15}[tl]{\footnotesize0.15}
\psfrag{0.20}[tl]{\footnotesize0.20}
\psfrag{0.000}[]{\footnotesize0}
\psfrag{0.005}[]{\footnotesize0.05}
\psfrag{0.010}[]{\footnotesize0.10}
\psfrag{0.015}[]{\footnotesize0.15}
\psfrag{0.020}[]{\footnotesize0.20}
\psfrag{0.025}[]{\footnotesize0.25}
\begin{center}
\includegraphics[width=.56\textwidth]{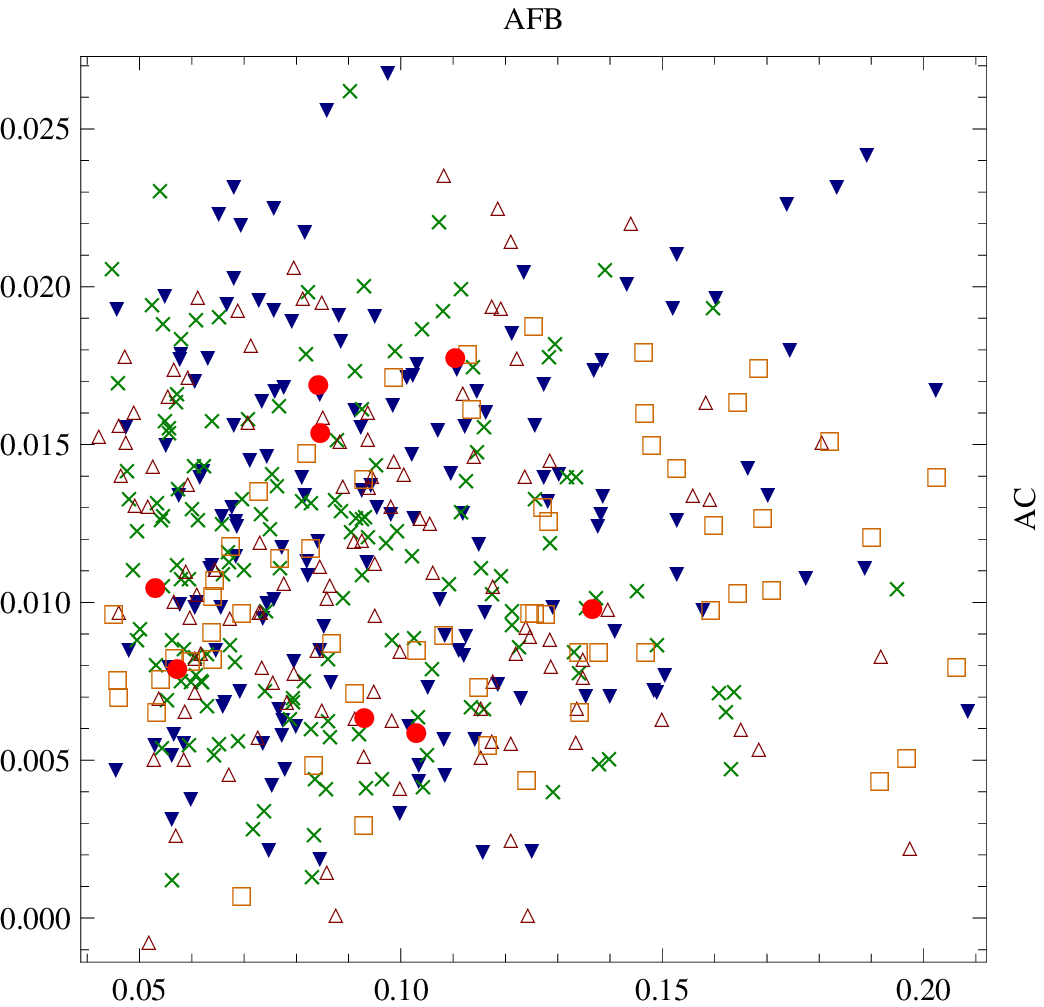}
\newline
\vskip0.01cm
\hspace*{-1.2cm}  (b)
\end{center}
\end{minipage}
\vspace*{0.1cm}
\caption{Predictions for LHC at 7 TeV for all the points of the parameter space
that reproduce the Tevatron measurements using the random scan of
section \ref{numeric}. Fig.~(a) shows the production
cross section of $G^*$ decaying to $jj$ and $t\bar t$. In one axis we
plot the LO production cross section of $G^*$ decaying to jets,
without b-tagging, after applying the set of kinematic cuts
described in \cite{updatedijet}. On the other axis we plot
$\sigma_{G^*}\times BR(G^*\to t\bar t)$ at LO, at the parton-level.
Fig.~(b) shows the total $A_{FB}^t$ at Tevatron versus total $A_C$ at
LHC at 7 TeV. The colours and style codify the different values for the
resonance mass: red dots for $M=700$ GeV, orange squares for $M=850$
GeV, brown empty triangles for
$M=1$ TeV, green crosses for $M=1.5$ TeV, and blue filled triangles
for $M=2.5$ TeV. The
corresponding coloured lines show the $95\%$ confidence-level upper
limit on dijet and top-pair production, taken
from~\cite{ttbarAtlas200} and~\cite{updatedijet}, from right to left
they correspond to increasing resonance masses. None horizontal blue
line has been drawn because there are no experimental limits on
$t\bar t$ production for a gluon resonance with $M=2.5$ TeV.}
\label{figjj-tt}
\end{figure}

We summarize the predictions and constraints from resonances searches
in $jj$ and $t\bar t$ creation at LHC in Fig.~\ref{figjj-tt}a. The
different colours codify the different values of the resonance mass:
red for $M=700$ GeV, orange for $M=850$ GeV, brown for $M=1$ TeV,
green for $M=1.5$ TeV, and blue for $M=2.5$ TeV. We also show with
coloured lines the upper bounds for each mass in the contribution to
dijets and top-pair production at $95\%$ confidence-level, taken from
Fig.~2 of~\cite{updatedijet} and Fig.~6 of~\cite{ttbarAtlas200}. There
is no upper bound line on $t\bar t$ production for $M=2.5$ TeV because
Ref.~\cite{ttbarAtlas200} only provides information up to $M=1.6$ TeV.
Only the points on the lower-left corner of these lines are not ruled
out. Note that although the limits from $t\bar t$ do not exclude any
point, a mild increase of the accumulated luminosity may be enough to
test a resonance with $M\gtrsim 1$ TeV. For $M\lesssim 1.5$ TeV the
dijet constraints are much above from the predictions of the model,
whereas for $M=2.5$ TeV approximately $88\%$ of the points are
excluded by dijet constraints. In this case only points with
$f_q\lesssim 1$ are allowed, therefore the non-observation of any
resonance in dijet final states selects the smaller couplings.

Recently, the CMS Collaboration has reported the first measurement of
the charge asymmetry in top quark pair production at LHC, with an
integrated luminosity of 36 pb$^{-1}$ at 7 TeV. The measured asymmetry
$A_C=0.060\pm0.134(\rm{stat.})\pm 0.026(\rm{syst.})$, being still
dominated by the statistical uncertainties \cite{AC-CMS}, is
consistent with the SM prediction $0.0130\pm0.0011$. Increasing the
integrated luminosity to $\gtrsim 1$ fb$^{-1}$, CMS can reach the same
sensitivity as Tevatron results, providing an independent measurement
of top asymmetries that can complement the Tevatron results on
$A_{FB}^t$. Moreover, as shown in Ref.~\cite{AguilarSaavedra:2011hz}, there is a correlation between both asymmetries, that depends on the
details of NP and can thus discriminate between different models. The
charge asymmetry measured by CMS is defined by:
\begin{equation}
A_C=\frac{N^+-N^-}{N^++N^-} \ ,
\end{equation}
with $N^\pm$ the number of events with positive or negative values of
$|\eta_t|-|\eta_{\bar t}|$, being $\eta$ the pseudo-rapidity in the
laboratory frame, $\eta=-\log\tan\theta/2$.

We have computed the charge asymmetry $A_C$ for top-pair production at
LHC at 7 TeV. The NP contribution to the total charge asymmetry in
$t\bar t$ production, $A_C^{(NP)}$, is predicted to be positive and of
order $0.01-0.03$ in our model, below the sensitivity of the present
measurements. Although there is a correlation with $A_{FB}^t$, it is
different from the one reported for the axial model in
\cite{AguilarSaavedra:2011hz}, essentially because the couplings are
neither axial, nor universal, resulting in a more complicated pattern
that is difficult to disentangle from the signals of other models by
just analyzing these observables: $A_{FB}^t$ and $A_C$. In
Fig.~\ref{figjj-tt}b we show our results for the contributions to
$A_{FB}^t$ and $A_C$. The different colours codify the resonance mass
as explained before.

In~\cite{AC-CMS} the CMS Collaboration has also provided the unfolded
results on the spectrum of $|\eta_t|-|\eta_{\bar t}|$, considering 6
bins, defined by the following limits for $|\eta_t|-|\eta_{\bar t}|$:
$-1, -0.5, 0, 0.5, 1$. Thus we have computed the spectrum of
$|\eta_t|-|\eta_{\bar t}|$, obtaining deviations of order $0.1-5\%$
compared with the SM prediction for these 6 bins. We obtain the
largest deviations in the bin $|\eta_t|-|\eta_{\bar t}|>1$, with
positive NP contributions of order $1-5\%$. However, since the
uncertainties are still large, all the points we have simulated are in
agreement with the data. Either larger luminosity or a more refined
resolution is needed to test our model with the $|\eta_t|-|\eta_{\bar
t}|$ spectrum.

There is also an interesting prediction for LHC concerning the $t\bar
t$ invariant mass spectrum. For all the points that agree with the
Tevatron data, our model results in an excess of events for $M_{t\bar
t}\lesssim (M-\Gamma_{G^*})$ and a lack of events for $M_{t\bar
t}\gtrsim (M+\Gamma_{G^*})$, compared with the SM. This can be seen
from the parton-level analysis of section~\ref{analytic}. Therefore,
detailed data on the $M_{t\bar t}$-spectrum could provide a strong
test for this kind of gluon resonance.

\section{Discussion and UV completions}\label{models}
We have considered an effective approach with just one gluon-resonance. There are many UV completions of the SM that contain a massive colour-octet vector as $G^*$, as for example composite Higgs models and extra-dimensions. In these theories there are in general resonances associated to the other SM fields also, and even infinite towers of massive resonances. We have considered just the effects of the lightest resonance associated to the gluon for simplicity, and because in many cases it gives the leading contributions to the physical processes that we have studied in this work. Examples of UV completions of our model have been presented recently in Refs.~\cite{Djouadi:2011aj} and~\cite{granada}, corresponding to two different models with a warped extra dimension, the main difference between them being the resonance scale, $1.5-2$ TeV and $\lesssim1$ TeV, respectively. Our results are compatible with the analysis of those papers, but the region of the parameter space that we explore is much larger than the one allowed in those frameworks. Usually, in warped-extra dimensional models, the hierarchy in the fermionic spectrum is explained by the localization of the 0-mode Kaluza-Klein (KK) wave functions in the extra dimension. These wave functions also determine the strength of the couplings with the massive KK modes, as $G^*$. In our model we have considered random couplings, without that kind of constraints. As explained in~\cite{Alvarez:2010js}, in a large family of composite Higgs models and in simple extra dimensional models it is not possible to obtain large axial couplings, because $f_\psi$ is either positive or small and negative. For this reason our results go beyond this kind of models. Interestingly, the results on $A_{FB}^t$ choose a region with large top-couplings, characteristic of models with a composite top.

We have shown that, to reproduce $A_{FB}^t$, the couplings of the light quarks can not be too small for $M\gtrsim 2$ TeV. For warped extra-dimensions, this implies that at least one of the chiralities of the light quarks has to be somewhat delocalized from the UV boundary, leading to sizable couplings not only with $G^*$, but also with the whole set of resonances associated to the EW sector. The precision EW measurements put strong constraints on these couplings, that usually can only be satisfied by introducing new symmetries~\cite{Agashe:2003zs,Agashe:2006at} or more structure~\cite{Delaunay:2011vv,Cabrer:2010si}. In Refs.~\cite{DaRold:2010as,Alvarez:2010js} we have shown how, by properly choosing the quantum numbers of the fermions under the custodial symmetry and an extra discrete symmetry (called $P_{LR}$ and $P_{C}$), one can protect the $Z$-couplings of the light fermions and reproduce the measured $A_{FB}^b$.\footnote{See also Refs. \cite{Djouadi:2006rk,Bouchart:2008vp,Djouadi:2011aj} for other solutions.} The same mechanism can be implemented in the present set-up for $M\gtrsim2$ TeV.

\section{Conclusions}
\label{conclusions}

In this work we have studied the phenomenology at Tevatron and LHC of a
TeV colour-octet vector resonance. We have focused our attention on
the top-pair creation at Tevatron, making an analytical study of the
new physics contribution to the cross section and forward-backward
asymmetry as functions of $M_{t\bar t}$ for different configurations.
This analysis shows the interplay between the interference and the new
physics square terms depending on the size of the couplings and the
resonance mass. It allows us to select the regions of the parameter
space that can increase the asymmetry in the high invariant mass
region without spoiling the cross section. We have also simulated the
top-pair creation at Tevatron at LO and at partonic level with {\tt Madgraph/Madevent}.
The previous analysis also allows us to understand the results of our
simulations. Finally, we have computed the contributions of our model
to dijet final states and top-pair creation at LHC, contrasting our
results with the limits on the production of resonances. We have also
shown the LHC predictions of our model for the charge asymmetry in
$t\bar t$ creation.

As a summary of our results for Tevatron, we have shown that an
$s$-channel gluon resonance with a mass $M=0.7-2.5$ TeV and chiral
couplings with the SM quarks can reproduce the data on top-pair
production with $p>0.05$, alleviating the tension present in the high
mass regime of $A^t_{FB}$, that exceeds the SM prediction by more than
three standard deviations. Almost all the previous analysis have
focused either on effective operators or heavy resonances to keep the
agreement with the SM cross section: $M\gtrsim1.5$ TeV (see footnote
1), whereas we have shown that resonant new
physics can also do the job. We have determined the couplings that
reproduce the data for different values of the resonance mass:
$M=700,\ 850,\ 1000,\ 1500$ and 2500 GeV, demanding small couplings
for the light quarks to avoid spoiling the LHC jet phenomenology. We
have found that in general axial-like couplings are preferred and
that, since we have chosen small light-quark couplings, the top
couplings are large $\gtrsim 1$, pointing towards top compositeness.
The size of the couplings of the light-quarks increases as the
resonance becomes heavier to keep a sizable contribution to the
asymmetry, with $f_q\sim 0.1$ for $M=700$ GeV and $f_q\sim 1$ for
$M=2.5$ TeV. We have also found that the density of points in the
parameter space that can accommodate the data is much smaller for
$M=700$ GeV than for $M\gtrsim 1$ TeV.

Concerning the LHC phenomenology, we have found that for $M=700,\
850,\ 1000$ and 1500 GeV, the limits on the search for new resonances
are above the predictions of our model. However, for $M=2.5$ TeV,
$88\%$ of the parameter space is ruled out by the limits on new
resonances set by dijets, remaining only the points with light-quark
couplings not larger than $\sim 1$. Our results also show that a large
region of the parameter space that lies below the present limits could
be tested with a sensible rise of the accumulated luminosity and the
center of mass energy.   It will be interesting to have experimental data on dijet resonance searches with $b$-tagging in order to discriminate regions in the parameter space.  
We have also computed the top charge asymmetry,
obtaining new physics contributions of order $1-3\%$, and corrections
to the $|\eta_t|-|\eta_{\bar t}|$ spectrum.

Note that, if the Tevatron data is explained by a gluon resonance, we
have to demand $M\gtrsim 700$ GeV, with light-quark couplings
increasing with the resonance mass. On the other hand, the limits from
LHC resonance searches seem to avoid $M\gtrsim 2.5$ TeV if the
couplings are large. Therefore, our analysis suggests that there is a
preferred window for the resonance mass: $M\sim0.7-2.5$ TeV.
Interestingly, a resonance like this can be tested at LHC in the next
years.

Ref.~\cite{CDF3-Asym} has provided a distribution of the events and of
the asymmetry as functions of the invariant mass, with bins of 15 GeV 
for the cross section and 50-100 GeV for the asymmetry.
However these distributions are not unfolded, making it difficult to
test the models beyond the SM with this level of accuracy. Notice that
the parton-level unfolded data presented for the asymmetry is divided
in two bins only~\cite{CDF3-Asym}. A more refined distribution at the
parton-level would be very useful to better test the new physics.

As a final consideration, note that for certain 
regions of the parameter space, our model can be embedded
in a more complete theory solving the hierarchy problem, as composite
Higgs models or extra dimensions. However there are some combinations
of couplings that either can not be obtained in these models, or fail
to reproduce the EW precision data when the full EW sector is
considered. It would be very interesting to find a model able to cover
those regions of the parameter space.
\section*{Acknowledgments}
We thank Jos\'e Santiago for a careful reading of the manuscript and suggestions. We also thank Carlos Wagner, Ricardo Piegaia, Mar\'{\i}a Teresa Dova, Xavier Anduaga and Cedric Delaunay for useful conversations. This work has been partially supported by ANPCyT (Argentina) under grant \# PICT-PRH 2009-0054 and by CONICET (Argentina) PIP-2011.
{}


\begin{thebibliography}{}

\bibitem{topNP}
  C.~T.~Hill, S.~J.~Parke,
  Phys.\ Rev.\  {\bf D49 } (1994)  4454-4462.
  [arXiv:hep-ph/9312324 [hep-ph]];
  C.~T.~Hill, E.~H.~Simmons,
  Phys.\ Rept.\  {\bf 381 } (2003)  235-402.
  [hep-ph/0203079];
  K.~Agashe, R.~Contino, R.~Sundrum,
  Phys.\ Rev.\ Lett.\  {\bf 95 } (2005)  171804.
  [hep-ph/0502222].

\bibitem{CDF1-Asym} CDF Public Note: http://www-cdf.fnal.gov/physics/new/top/2010/tprop\\/Afb/cdfnote\_10224\_public\_v02\_for\_10185\_ttbarAfbDeltay.pdf

\bibitem{CDF2-Asym} T.~Aaltonen {\it et al.} [ CDF Collaboration ], 
Phys.\ Rev.\ Lett.\  {\bf 101}, 202001 (2008) [arXiv:0806.2472 [hep-ex]].

\bibitem{D01-Asym} V.~M.~Abazov {\it et al.} [ D0 Collaboration ], 
Phys.\ Rev.\ Lett.\  {\bf 100}, 142002 (2008) [arXiv:0712.0851 [hep-ex]].

\bibitem{ttbar} 
  G.~Burdman, L.~de Lima, R.~D.~Matheus,
  Phys.\ Rev.\  {\bf D83 } (2011)  035012.
  [arXiv:1011.6380 [hep-ph]];
  B.~Xiao, Y.~-k.~Wang, S.~-h.~Zhu,
  [arXiv:1011.0152 [hep-ph]];
  C.~Degrande, J.~-M.~Gerard, C.~Grojean, F.~Maltoni, G.~Servant,
  JHEP {\bf 1103 } (2011)  125.
  [arXiv:1010.6304 [hep-ph]];
  C.~-H.~Chen, G.~Cvetic, C.~S.~Kim,
  PHLTA,B694,393-397.\ 2011 {\bf B694 } (2011)  393-397.
  [arXiv:1009.4165 [hep-ph]];
  J.~A.~Aguilar-Saavedra,
  Nucl.\ Phys.\  {\bf B843 } (2011)  638-672.
  [arXiv:1008.3562 [hep-ph]];
  M.~Bauer, F.~Goertz, U.~Haisch, T.~Pfoh, S.~Westhoff,
  JHEP {\bf 1011 } (2010)  039.
  [arXiv:1008.0742 [hep-ph]];
  R.~S.~Chivukula, E.~H.~Simmons, C.~-P.~Yuan,
  Phys.\ Rev.\  {\bf D82 } (2010)  094009.
  [arXiv:1007.0260 [hep-ph]];
  B.~Xiao, Y.~-k.~Wang, S.~-h.~Zhu,
  Phys.\ Rev.\  {\bf D82 } (2010)  034026.
  [arXiv:1006.2510 [hep-ph]];
  Q.~-H.~Cao, D.~McKeen, J.~L.~Rosner, G.~Shaughnessy, C.~E.~M.~Wagner,
  Phys.\ Rev.\  {\bf D81 } (2010)  114004;
  [arXiv:1003.3461 [hep-ph]];
  J.~Cao, Z.~Heng, L.~Wu, J.~M.~Yang,
  Phys.\ Rev.\  {\bf D81 } (2010)  014016.
  [arXiv:0912.1447 [hep-ph]];
  I.~Dorsner, S.~Fajfer, J.~F.~Kamenik, N.~Kosnik,
  Phys.\ Rev.\  {\bf D81 } (2010)  055009.
  [arXiv:0912.0972 [hep-ph]];
  A.~Arhrib, R.~Benbrik, C.~-H.~Chen,
  Phys.\ Rev.\  {\bf D82 } (2010)  034034.
  [arXiv:0911.4875 [hep-ph]];
  J.~Shu, T.~M.~P.~Tait, K.~Wang,
  Phys.\ Rev.\  {\bf D81 } (2010)  034012.
  [arXiv:0911.3237 [hep-ph]];
  P.~H.~Frampton, J.~Shu, K.~Wang,
  Phys.\ Lett.\  {\bf B683 } (2010)  294-297.
  [arXiv:0911.2955 [hep-ph]];
  K.~Cheung, W.~-Y.~Keung, T.~-C.~Yuan,
  Phys.\ Lett.\  {\bf B682 } (2009)  287-290.
  [arXiv:0908.2589 [hep-ph]];
  S.~Jung, H.~Murayama, A.~Pierce, J.~D.~Wells,
  Phys.\ Rev.\  {\bf D81 } (2010)  015004.
  [arXiv:0907.4112 [hep-ph]];
  P.~Ferrario, G.~Rodrigo,
  Phys.\ Rev.\  {\bf D80 } (2009)  051701.
  [arXiv:0906.5541 [hep-ph]];
  M.~V.~Martynov, A.~D.~Smirnov,
  Mod.\ Phys.\ Lett.\  {\bf A24 } (2009)  1897-1905.
  [arXiv:0906.4525 [hep-ph]];
  K.~Kumar, T.~M.~P.~Tait, R.~Vega-Morales,
  JHEP {\bf 0905 } (2009)  022.
  [arXiv:0901.3808 [hep-ph]]; and Ref.~\cite{Alvarez:2010js,Djouadi:2009nb}.

\bibitem{CDF3-Asym}
  T.~Aaltonen {\it et al.} [ CDF Collaboration ],
  [arXiv:1101.0034 [hep-ex]].

\bibitem{ttbar2}
  E.~Gabrielli, M.~Raidal,
  [arXiv:1106.4553 [hep-ph]];
  Y.~Cui, Z.~Han, M.~D.~Schwartz,
  [arXiv:1106.3086 [hep-ph]];
  J.~A.~Aguilar-Saavedra, M.~Perez-Victoria,
  [arXiv:1105.4606 [hep-ph]];
M.~Duraisamy, A.~Rashed and A.~Datta,
 [arXiv:1106.5982 [hep-ph]];
  D.~Krohn, T.~Liu, J.~Shelton, L.~-T.~Wang,
  [arXiv:1105.3743 [hep-ph]];
  K.~S.~Babu, M.~Frank, S.~K.~Rai,
  [arXiv:1104.4782 [hep-ph]];
  D.~-W.~Jung, P.~Ko, J.~S.~Lee,
  [arXiv:1104.4443 [hep-ph]];
  S.~Jung, A.~Pierce, J.~D.~Wells,
  [arXiv:1104.3139 [hep-ph]];
  C.~Degrande, J.~-M.~Gerard, C.~Grojean, F.~Maltoni, G.~Servant,
  [arXiv:1104.1798 [hep-ph]];
  X.~-P.~Wang, Y.~-K.~Wang, B.~Xiao, J.~Xu, S.~-h.~Zhu,
  PHRVA,D83,115010.\ 2011 {\bf D83 } (2011)  115010.
  [arXiv:1104.1917 [hep-ph]];
  A.~E.~Nelson, T.~Okui, T.~S.~Roy,
  [arXiv:1104.2030 [hep-ph]];
  J.~A.~Aguilar-Saavedra, M.~Perez-Victoria,
  PHLTA,B701,93-100.\ 2011 {\bf B701 } (2011)  93-100.
  [arXiv:1104.1385 [hep-ph]];
  A.~Rajaraman, Z.~'e.~Surujon, T.~M.~P.~Tait,
  [arXiv:1104.0947 [hep-ph]];
  J.~Shu, K.~Wang, G.~Zhu,
  [arXiv:1104.0083 [hep-ph]];
  S.~Jung, A.~Pierce, J.~D.~Wells,
  [arXiv:1103.4835 [hep-ph]];
  M.~I.~Gresham, I.~-W.~Kim, K.~M.~Zurek,
  [arXiv:1103.3501 [hep-ph]];
  Z.~Ligeti, M.~Schmaltz, G.~M.~Tavares,
  JHEPA,1106,109.\ 2011 {\bf 1106 } (2011)  109.
  [arXiv:1103.2757 [hep-ph]];
  C.~Grojean, E.~Salvioni, R.~Torre,
  [arXiv:1103.2761 [hep-ph]];
  J.~A.~Aguilar-Saavedra, M.~Perez-Victoria,
  JHEP {\bf 1105 } (2011)  034.
  [arXiv:1103.2765 [hep-ph]];
  C.~Delaunay, O.~Gedalia, Y.~Hochberg, G.~Perez, Y.~Soreq,
  [arXiv:1103.2297 [hep-ph]];
  R.~Foot,
  Phys.\ Rev.\  {\bf D83 } (2011)  114013.
  [arXiv:1103.1940 [hep-ph]];
  N.~Craig, C.~Kilic, M.~J.~Strassler,
  [arXiv:1103.2127 [hep-ph]];
  E.~R.~Barreto, Y.~A.~Coutinho, J.~Sa Borges,
  Phys.\ Rev.\  {\bf D83 } (2011)  054006.
  [arXiv:1103.1266 [hep-ph]];
  A.~R.~Zerwekh,
  [arXiv:1103.0956 [hep-ph]];
  G.~Isidori, J.~F.~Kamenik,
  Phys.\ Lett.\  {\bf B700 } (2011)  145-149.
  [arXiv:1103.0016 [hep-ph]];
  K.~M.~Patel, P.~Sharma,
  JHEP {\bf 1104 } (2011)  085.
  [arXiv:1102.4736 [hep-ph]];
  B.~Grinstein, A.~L.~Kagan, M.~Trott, J.~Zupan,
  PRLTA,107,012002.\ 2011 {\bf 107 } (2011)  012002.
  [arXiv:1102.3374 [hep-ph]];
  V.~Barger, W.~-Y.~Keung, C.~-T.~Yu,
  Phys.\ Lett.\  {\bf B698 } (2011)  243-250.
  [arXiv:1102.0279 [hep-ph]];
  M.~I.~Gresham, I.~-W.~Kim, K.~M.~Zurek,
  [arXiv:1102.0018 [hep-ph]];
  Y.~Bai, J.~L.~Hewett, J.~Kaplan, T.~G.~Rizzo,
  JHEP {\bf 1103 } (2011)  003.
  [arXiv:1101.5203 [hep-ph]];
  K.~Cheung, T.~-C.~Yuan,
  Phys.\ Rev.\  {\bf D83 } (2011)  074006.
  [arXiv:1101.1445 [hep-ph]]; and Refs.~\cite{granada,Djouadi:2011aj,Barcelo:2011vk}.

\bibitem{talkPlanck2011} 
  E.~Alvarez, L.~Da Rold, J.I.~Sanchez Vietto and A.~Szynkman, ``Weakly coupled resonant NP at the Tevatron t-tbar forward-backward asymmetry,'' presented at PLANCK 2011 - From the Planck Scale to the ElectroWeak Scale, Lisbon, May 30th to June 3rd (2011).

\bibitem{granada}
  R.~Barcelo, A.~Carmona, M.~Masip, J.~Santiago,
   [arXiv:1105.3333 [hep-ph]].

\bibitem{gaugeKK}
  H.~Davoudiasl, J.~L.~Hewett and T.~G.~Rizzo,
  Phys.\ Lett.\  B {\bf 473} (2000) 43
  [arXiv:hep-ph/9911262];
  A.~Pomarol,
  Phys.\ Lett.\  B {\bf 486} (2000) 153
  [arXiv:hep-ph/9911294].


\bibitem{Contino:2006nn}
  R.~Contino, T.~Kramer, M.~Son, R.~Sundrum,
  JHEP {\bf 0705 } (2007)  074.
  [hep-ph/0612180].

\bibitem{DaRold:2010as}
  L.~Da Rold,
  JHEP {\bf 1102} (2011) 034
  [arXiv:1009.2392 [hep-ph]].

\bibitem{Alvarez:2010js}
  E.~Alvarez, L.~Da Rold and A.~Szynkman,
  JHEP {\bf 1105} (2011) 070
  [arXiv:1011.6557 [hep-ph]].

\bibitem{Djouadi:2009nb}
  A.~Djouadi, G.~Moreau, F.~Richard and R.~K.~Singh,
  Phys.\ Rev.\  D {\bf 82} (2010) 071702
  [arXiv:0906.0604 [hep-ph]].

\bibitem{Djouadi:2006rk}
  A.~Djouadi, G.~Moreau and F.~Richard,
  Nucl.\ Phys.\  B {\bf 773} (2007) 43
  [arXiv:hep-ph/0610173].

\bibitem{Djouadi:2011aj}
  A.~Djouadi, G.~Moreau and F.~Richard,
  arXiv:1105.3158 [hep-ph].

\bibitem{Barcelo:2011vk}
  R.~Barcelo, A.~Carmona, M.~Masip, J.~Santiago,
  [arXiv:1106.4054 [hep-ph]].

\bibitem{mgme}
  J.~Alwall, P.~Demin, S.~de Visscher, R.~Frederix, M.~Herquet, F.~Maltoni, T.~Plehn, D.~L.~Rainwater {\it et al.},
  JHEP {\bf 0709 } (2007)  028,
  [arXiv:0706.2334 [hep-ph]].

\bibitem{Pumplin:2002vw}
  J.~Pumplin, D.~R.~Stump, J.~Huston, H.~L.~Lai, P.~M.~Nadolsky and W.~K.~Tung,
  JHEP {\bf 0207} (2002) 012
  [arXiv:hep-ph/0201195].
 
\bibitem{b-precision} 
  S.~Chang, C.~S.~Kim, J.~Song,
  JHEP {\bf 0702 } (2007)  087.
  [hep-ph/0607313]; 
  C.~Csaki, D.~Curtin,
  Phys.\ Rev.\  {\bf D80 } (2009)  015027.
  [arXiv:0904.2137 [hep-ph]].


%
%
%

\bibitem{ttbarAtlas35}
ATLAS-CONF-2011-023,
the Single Lepton+Jets Channel without b-tagging'' [ATLAS
collaboration];
ATLAS-CONF-2011-034, 
cross section with ATLAS in pp collisions at sqrt(s) = 7 TeV in
dilepton final states'' [ATLAS collaboration];
ATLAS-CONF-2011-035, 
cross section with ATLAS in pp collisions at sqrt(s) = 7 TeV in the
single-lepton channel using b-tagging'' [ATLAS collaboration];
ATLAS-CONF-2011-040, 
production cross section using dilepton and single-lepton final
states'' [ATLAS collaboration];
ATLAS-CONF-2011-066, 
all-hadronic channel in ATLAS with $\sqrt{s}$ = 7 TeV data''
[ATLAS collaboration].

\bibitem{ttbarAtlas200}
ATLAS-CONF-2011-087, ``A Search for $t\bar t$ Resonances in the Lepton
Plus Jets Channel using 200 pb$^{-1}$ of $pp$ Collisions at $\sqrt{s}$=7 TeV'', [ATLAS collaboration].

\bibitem{Aad:2011aj}
 G.~Aad {\it et al.} [ ATLAS Collaboration ],
 New J.\ Phys.\  {\bf 13 } (2011)  053044.
 [arXiv:1103.3864 [hep-ex]].

\bibitem{updatedijet}
ATLAS-CONF-2011-081, ``Update of the Search for New Physics in the Dijet Mass Distribution in 163 pb$^{-1}$ of pp Collisions at $\sqrt{s}$=7 TeV Measured with the ATLAS Detector,'' [ATLAS collaboration]. 

\bibitem{Sjostrand:2006za}
  T.~Sjostrand, S.~Mrenna and P.~Z.~Skands,
  JHEP {\bf 0605} (2006) 026
  [arXiv:hep-ph/0603175].






\bibitem{AC-CMS}
CMS-PAS-TOP-10-010, ``Measurement of the charge asymmetry in top quark
pair production, with the CMS experiment''.

\bibitem{AguilarSaavedra:2011hz}
 J.~A.~Aguilar-Saavedra and M.~Perez-Victoria,
 arXiv:1105.4606 [hep-ph].

\bibitem{Agashe:2003zs}
  K.~Agashe, A.~Delgado, M.~J.~May, R.~Sundrum,
  JHEP {\bf 0308 } (2003)  050,
  [arXiv:hep-ph/0308036].

\bibitem{Agashe:2006at}
  K.~Agashe, R.~Contino, L.~Da Rold, A.~Pomarol,
  Phys.\ Lett.\  {\bf B641 } (2006)  62-66,
  [arXiv:hep-ph/0605341].

\bibitem{Delaunay:2011vv}
  C.~Delaunay, O.~Gedalia, S.~J.~Lee, G.~Perez, E.~Ponton,
  [arXiv:1101.2902 [hep-ph]].

\bibitem{Cabrer:2010si}
  J.~A.~Cabrer, G.~von Gersdorff, M.~Quiros,
  Phys.\ Lett.\  {\bf B697 } (2011)  208-214,
  [arXiv:1011.2205 [hep-ph]],
  JHEP {\bf 1105 } (2011)  083.
  [arXiv:1103.1388 [hep-ph]].

\bibitem{Bouchart:2008vp}
  C.~Bouchart, G.~Moreau,
  Nucl.\ Phys.\  {\bf B810 } (2009)  66-96.
  [arXiv:0807.4461 [hep-ph]].

\end{thebibliography}
\end{document}